%% file: main.tex
\documentclass[preprint,a4paper,notoc]{JHEP3}
\usepackage{amsmath}
\usepackage{epsfig}
\usepackage{multicol}
\usepackage{dsfont}
\usepackage{psfrag}
\usepackage{amssymb}
\usepackage{bbm}
\usepackage{cite}
\newcommand{\tr}{\mbox{tr}}

\title{A strategy for implementing non-perturbative renormalisation of
heavy-light four-quark operators in the static approximation}

\author{\epsfig{figure=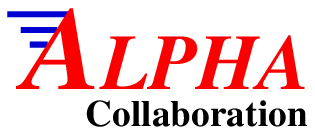,width=2.5cm}}

\author{
  Filippo Palombi\\
  DESY, Theory Group, Notkestra\ss e 85, D-22603 Hamburg, Germany\\
  E-mail: \email{filippo.palombi@desy.de}}
\author{
  Mauro Papinutto\\
  INFN Sez. di Roma 3,
  Via della Vasca Navale 84, 00146 Rome, Italy\\
  E-mail: \email{mauro.papinutto@fis.uniroma3.it}}
\author{
  Carlos Pena\\
  CERN, Physics Department, Theory Division, CH-1211 Geneva 23,
  Switzerland\\
  E-mail: \email{carlos.pena.ruano@cern.ch}}
\author{
  Hartmut Wittig\\
  Institut f\"ur Kernphysik, University of Mainz, D-55099 Mainz, Germany\\
  E-mail: \email{wittig@kph.uni-mainz.de}}

\preprint{DESY 06-030\\ RM3-TH/06-5\\ CERN-PH-TH/2006-043\\ MKPH-T-0605}

\abstract{We discuss the renormalisation properties of the complete
  set of $\Delta{B}=2$ four-quark operators with the heavy quark
  treated in the static approximation. We elucidate the r\^ole of
  heavy quark symmetry and other symmetry transformations in
  constraining their mixing under renormalisation. By employing the
  Schr\"odinger functional, a set of non-perturbative renormalisation
  conditions can be defined in terms of suitable correlation
  functions. As a first step in a fully non-perturbative determination
  of the scale-dependent renormalisation factors, we evaluate these
  conditions in lattice perturbation theory at one loop. Thereby we
  verify the expected mixing patterns and determine the anomalous
  dimensions of the operators at NLO in the Schr\"odinger functional
  scheme. Finally, by employing twisted-mass QCD it is shown how
  finite subtractions arising from explicit chiral symmetry breaking
  can be avoided completely.}


\usepackage[OT2,OT1]{fontenc}
\newcommand\cyr{%
\renewcommand\rmdefault{wncyr}%
\renewcommand\sfdefault{wncyss}%
\renewcommand\encodingdefault{OT2}%
\normalfont
\selectfont}
\DeclareTextFontCommand{\textcyr}{\cyr} 

\newcommand{\bx}{{\mathbf{x}}}

\newcommand{\cF}{{\cal F}}
\newcommand{\cO}{{\cal O}}
\newcommand{\cQ}{{\cal Q}}
\newcommand{\cR}{{\cal R}}
\newcommand{\cS}{{\cal S}}
\newcommand{\cX}{{\cal X}}
\newcommand{\cW}{{\cal W}}
\newcommand{\cZ}{{\cal Z}}
\newcommand{\dZ}{{\mathbb{Z}}}
\newcommand{\fla}{s}

\newcommand{\vx}{\mathbf{x}}
\newcommand{\vy}{\mathbf{y}}

\newcommand{\pvs}{\cS}
\newcommand{\pvo}{\cQ}
\newcommand{\NC}{N}
\newcommand{\NF}{N_{\rm\scriptsize f}}
\newcommand{\Nf}{N_{\rm\scriptsize f}}
\newcommand{\eps}{\varepsilon}
\newcommand{\half}{{\scriptstyle{{1\over 2}}}}

\newcommand\Cf{C_{\rm F}}
\newcommand\csw{c_{\rm sw}}
\newcommand{\Oa}{\mbox{O}(a)}

\newcommand{\MSbar}{{$\overline{\rm MS}$\ }}
\newcommand{\Dslash}{\relax{\kern+.25em / \kern-.70em D}}
\newcommand{\lp}{{\scriptsize +}}
\newcommand{\lm}{{\scriptsize -}}
\newcommand{\eq}[1]{eq.~(\ref{#1})}

\newcommand{\pcd}{\Delta}
\newcommand{\pvd}{\mbox{\textcyr{D}}}

\begin{document}

\input intro

\input mixing

\input renorm_cond
\input perturb
\input mapping

\input conclusions

\input acknow
\input appendA

\input appendB
\input appendC

\input biblio

\end{document}

%% file: intro.tex
\section{Introduction}

The oscillations among pairs of neutral $B$-mesons provide crucial
information for pinning down the elements of the
Cabibbo-Kobayashi-Maskawa (CKM) Matrix that are associated with the
top quark. Owing to the absence of flavour-changing neutral currents
in the Standard Model, these oscillations are described by box
diagrams, in which the flavour change is brought about through the
intermediate propagation of a virtual top quark. By integrating out
the $W$-boson, the box diagram is replaced by an effective point-like
interaction vertex associated with the left-handed $\Delta{B}=2$
four-quark operator
\begin{align}
\label{OLL}
& \cO_{\rm LL}(x) = \bar b(x)\gamma_\mu^L\psi_\ell(x)\ \bar
   b(x)\gamma_\mu^L\psi_\ell(x)\ , 
\end{align}
where $\gamma_\mu^L=\gamma_\mu(\mathds{1}-\gamma_5)$, and the flavour
label $\ell$ denotes either a $d$ or an $s$ quark. The matrix elements
of $\cO_{\rm LL}$ between $B$-meson states are commonly parameterised
in terms of the $B$-parameters $B_B$ and $B_{B_s}$, for instance
\begin{align}
\label{B_B}
& \langle \bar B^0|\cO_{\rm LL}|B^0\rangle = \frac{8}{3}m_B^2f_B^2B_B\ ,
\end{align}
for $\ell=d$. The operator $\cO_{\rm LL}$ can be decomposed into
parity-even and parity-odd components $\cO_{\rm VV+AA}$ and $\cO_{\rm
VA+AV}$. In the Standard Model only the parity-even part $\cO_{\rm
VV+AA}$ makes a contribution to $B$-meson mixing. The $B$-parameters
encode the long-distance effects of the strong interaction and must be
determined in a non-perturbative approach such as lattice QCD. Indeed,
various lattice estimates of $B_B$ and $B_{B_s}$ have been published
by several authors in recent years
\cite{Abada:1991mt,Ewing:1995ih,Gimenez:1996sk,Christensen:1996sj,Bernard:1998dg,Gimenez:1998mw,Becirevic:2000nv,Hashimoto:2000eh,Lellouch:2000tw,Becirevic:2001xt,Aoki:2002bh,Aoki:2003xb}.
\\[1.5ex]
It is well known that relativistic $b$-quarks cannot be simulated
directly for currently accessible lattice spacings. Several formalisms
for treating $b$-quarks on the lattice, based on Heavy Quark Effective
Theory (HQET) \cite{Eichten:1987xu,Heitger:2003nj}, non-relativistic
QCD \cite{NRQCD}, on-shell improvement for relativistic quarks
\cite{El-Khadra:1996mp,Aoki:2001ra}, as well as finite-size scaling
techniques \cite{Guagnelli:2002jd}, have been developed and
applied. Obviously, some, if not all, of these approaches imply
certain approximations or assumptions whose validity and intrinsic
accuracy must be investigated.
\\[1.0ex]
In order to yield useful phenomenological information, operators like
$\cO_{\rm LL}$ must be renormalised. If the regulator breaks chiral
symmetry, as is the case for Wilson fermions, the renormalisation of
$\cO_{\rm LL}$, which has a particular chiral structure, is
complicated by the fact that it undergoes mixing with operators of
different chiralities. Therefore, in addition to an overall
logarithmically divergent, multiplicative renormalisation factor, one
must also determine finite subtraction coefficients.
\\[1.0ex]
The analogous case of $K^0-\bar K^0$ mixing, in which all fields that
appear in the corresponding four-quark operator are treated
relativistically, has been studied in ref. \cite{Donini:1999sf}. There
the renormalisation and mixing patterns of a general set of four-quark
operators were classified according to their transformation properties
under certain symmetries. In particular, it was shown how the mixing
due to explicit chiral symmetry breaking implied by the Wilson term
could be isolated and absorbed into mixing coefficients.  Another
important result of \cite{Donini:1999sf} was the observation that the
parity-odd component $\cO_{\rm VA+AV}$ is protected against mixing by
discrete symmetries.
\\[1.0ex]
In this paper we adopt a similar strategy to extend the analysis of
ref.~\cite{Donini:1999sf} to the case where the $b$-quark is treated
at leading order in HQET, i.e. in the static approximation. In
particular, we show how the heavy quark spin symmetry, in conjunction
with transformation properties under spatial rotations, as well as
discrete symmetries like parity and time reversal can be used to
constrain the renormalisation patterns of a general set of
static-light four-quark operators. One key result is that it is possible 
to find a basis of parity-odd operators that renormalise purely
multiplicatively. This allows to devise a strategy aimed at a
non-perturbative determination of the renormalisation factors
required for the calculation of $B$-parameters, without the need
to determine finite subtractions.
\\[1.0ex]
To this end we use twisted mass QCD (tmQCD)
\cite{Frezzotti:2000nk} as a discretisation for the light quark
fields, which allows us to map parity-even operators to parity-odd
ones. By employing the Schr\"odinger functional (SF)
\cite{Luscher:1992an}, the anomalous dimension of the latter can
then be determined non-perturbatively in complete analogy to the
case of $K^0-\bar K^0$ mixing studied previously in
\cite{Guagnelli:2005zc}. Thus, in order to reconstruct the
phenomenologically relevant matrix element of $\cO_{\rm LL}$, one
only needs to determine the renormalisation properties of
multiplicatively renormalisable operators, even in regularisations
that break chiral symmetry explicitely.
\\[1.0ex]
In addition to explaining how the renormalisation properties of
static-light four-quark operators can be constrained, another purpose
of this paper is to identify -- in the spirit of \cite{Palombi:2005zd}
-- suitable finite-volume renormalisation schemes based on
the SF, to be used in a forthcoming non-perturbative calculation. To
this end we have computed the anomalous dimension in perturbation
theory at NLO for the complete basis of four-quark operators in
several SF schemes.
\\[1.0ex]
Of course, the complicated mixing patterns one is confronted with when
using Wilson fermions can be avoided by using discretisations for the
light quarks which obey the Ginsparg-Wilson relation. First steps in
this direction have been taken in
ref. \cite{Becirevic:2003hd,Becirevic:2005sx}. However, in this work
we show, by using symmetry properties and tmQCD, that the
renormalisation of static-light four-quark operators describing
$B^0-\bar B^0$ mixing can be studied in an equally simple framework
for Wilson-type regularisations. Non-perturbative renormalisation can
thus be implemented in a straightforward manner and at much reduced
computational cost.
\\[1.0ex]
This paper is organised as follows: in section \ref{sec:mixing} we
discuss how the transformation properties under various symmetries
constrain the mixing patterns of static-light four-quark operators. In
section \ref{sec:ren_cond} we formulate a set of renormalisation
conditions for the operator basis within the Schr\"odinger
functional. Section \ref{sec:perturb} describes the perturbative
calculation which yields the NLO anomalous dimensions of the operators
for a set of Schr\"odinger functional renormalisation schemes. In
section \ref{sec:mapping} we discuss the use of tmQCD to compute the
physical matrix elements for $B^0$--$\bar B^0$ mixing using
multiplicatively renormalisable operators. Our
conclusions are presented in section \ref{sec:concl}. Technical
details regarding the use of symmetries to constrain the renormalisation
pattern and the evaluation of lattice integrals are relegated to Appendices
\ref{app:A} and \ref{app:B}, respectively. Tables listing the finite
parts of renormalisation constants and the NLO anomalous dimensions
can be found in Appendix~\ref{app:C}.

%% file: mixing.tex
\section{Mixing of heavy-light four-quark operators in the static
  approximation} 
\label{sec:mixing}

In this section we study the mixing of $\Delta B=2$ heavy-light
four-quark operators in which the heavy quarks are treated in the
static approximation of HQET. Thus they are represented by a pair of
static fields $(\psi_h,\psi_{\bar h})$, propagating forward and
backward in time, respectively; their dynamics is governed by the
Eichten-Hill action \cite{Eichten:1989zv}
\begin{gather}
\label{EHaction}
  S^{\rm stat}[\psi_h,\psi_{\bar h}] = a^4\sum_x
  \left[\bar\psi_h(x)\nabla_0^*\psi_h(x) -
    \bar\psi_{\bar h}(x)\nabla_0\psi_{\bar h}(x)\right],
\end{gather}
where the forward and backward covariant derivatives
$\nabla_0,\,\nabla_0^*$ are defined by
\begin{align}
& \nabla_0\psi_{\bar h}(x)=\frac{1}{a}\left[U_0(x)\psi_{\bar h}(x+a\hat{0})
 -\psi_{\bar h}(x)\right],  \nonumber \\
& \nabla_0^*\psi_h(x)=\frac{1}{a}\left[\psi_h(x)-U_0(x-a\hat{0})^{-1}
 \psi_h(x-a\hat{0})\right].
\label{eq_latt_der}
\end{align}
The field $\psi_h$($\bar \psi_{h}$) can be thought of as the
annihilator(creator) of a heavy quark. Similarly, $\psi_{\bar
  h}$($\bar\psi_{\bar h}$) creates(annihilates) a heavy 
antiquark. Each field is represented by a four-component Dirac vector,
yet only half of the components play a dynamical r\^ole, owing to the static 
projection constraints 
\begin{align}
\label{constraints}
& P_\lp\psi_h = \psi_h\ ,\qquad \bar\psi_hP_\lp = \bar\psi_h\ ,\qquad
  P_\lp = \frac{1}{2}(\mathds{1}+\gamma_0)\ ;  \nonumber \\
& P_\lm\psi_{\bar h} = \psi_{\bar h}\ ,\qquad \bar\psi_{\bar h}P_\lm =
  \bar\psi_{\bar h}\ ,\qquad P_\lm = \frac{1}{2}(\mathds{1}-\gamma_0)\ .
\end{align}
Instead of the link variables that appear in eq.\,(\ref{eq_latt_der})
one can consider more general definitions of the parallel transporter
which enters the covariant derivative. A set of alternative
discretisations was studied in \cite{DellaMorte:2005yc}, where it was
found that adequate choices of parallel transporter lead to much improved
signal-to-noise ratios in actual simulations.

The light (relativistic) quarks are instead taken to be Wilson
fermions, using either the plain Wilson action or its $\Oa$ improved
version with a Sheikholeslami-Wohlert
term~\cite{Sheikholeslami:1985ij}. The explicit chiral symmetry
breaking induced by the Wilson term causes the mixing of operator of
different naive chirality even in the chiral limit.

We consider a complete basis of $\Delta B = 2$ heavy-light four-quark 
operators which, for the sake of definiteness, we chose to contain two 
static fields $\bar\psi_{h}$ and $\bar\psi_{\bar h}$ 
while, in the light sector, we consider massless fermions with 
two distinct flavours $\psi_1$ and $\psi_2$. 
We introduce a generic $\Delta B = 2$ operator via
\begin{equation}
\label{genop}
\cO^\pm_{\Gamma_1\Gamma_2} =
    \dfrac{1}{2}\left[(\bar\psi_{h}\Gamma_1\psi_1)(\bar\psi_{\bar
    h}\Gamma_2\psi_2) \pm (\bar\psi_{h}\Gamma_1\psi_2)(\bar\psi_{\bar
    h}\Gamma_2\psi_1)\right]\ ,
\end{equation}
where $\Gamma_{1,2}$ are Dirac matrices, and we adopt the notation
\begin{equation}
\cO^\pm_{\Gamma_1\Gamma_2\ \pm\ \Gamma_3\Gamma_4} \equiv
\cO^\pm_{\Gamma_1\Gamma_2} \pm \cO^\pm_{\Gamma_3\Gamma_4}\ .
\end{equation}
The complete basis of Lorentz invariant operators is given by the set
of 16 operators
\begin{alignat}{3}
  \label{basis}
	\hbox{parity-even:~~} Q_1^\pm &= \cO^\pm_{\rm VV+AA}\ ,
 \qquad \hbox{parity-odd:~~} & {\cQ}_1^\pm &= \cO^\pm_{\rm VA+AV}\ ,
 \nonumber \\
	Q_2^\pm &= \cO^\pm_{\rm SS+PP}\ , & {\cQ}_2^\pm &= \cO^\pm_{\rm SP+PS}
\, \nonumber \\
	Q_3^\pm &= \cO^\pm_{\rm VV-AA}\ , & {\cQ}_3^\pm &= \cO^\pm_{\rm VA-AV}
\ , \nonumber \\
	Q_4^\pm &= \cO^\pm_{\rm SS-PP}\ , & {\cQ}_4^\pm &= \cO^\pm_{\rm SP-PS}
\ ,
\end{alignat}
which we have grouped according to their transformation properties
under parity. Here ${\rm V} = \gamma_\mu$, ${\rm A} =
\gamma_\mu\gamma_5$, $S = \mathds{1}$, ${\rm P}=\gamma_5$, and an
implicit summation over pairs of Lorentz indices is understood. We
incidentally remind the reader that tensor structures like
$T=\sigma_{\mu\nu}$ or $\tilde T=\sigma_{\mu\nu}\gamma_5$ produce
redundant operators in the static limit, due the projection
constraints (\ref{constraints}).

The description of $\Delta{B}=2$ transitions in terms of the static
approximation of HQET implies that the operator ${\cal{O}}_{\rm LL}$
of \eq{OLL} is related in some particular way to the operators listed
in \eq{basis}. Owing to the heavy-quark spin symmetry, one finds that
${\cal{O}}_{\rm LL}$ must be matched to a linear combination of
${\cal{O}}^{+}_{\rm VV+AA}$ and ${\cal{O}}^{+}_{\rm SS+PP}$
\cite{Flynn:1990qz}, and thus those two operators are of particular 
interest to the study of $B^0 - \bar B^0$ mixing.

The operator basis in eq.~(\ref{basis}) renormalises, in full
generality, via a $16 \times 16$ matrix $\dZ$, the form of which can
be constrained through symmetry arguments. A systematic method to
carry out this analysis is given by the following prescription:
\begin{itemize}
\item[(i)]{Construct the matrices $\Phi_k$ that implement, at the level
  of the operator basis, a maximal set of independent symmetry 
  transformations that leave the action invariant.}
\item[(ii)]{Impose the constraints
  \begin{gather}
  \dZ = \Phi_k\dZ\Phi_k^{-1},\
  \forall k. 
  \end{gather}
  The solution $\dZ$ to this system of equations displays
  the constrained form of the renormalisation matrix.}
\end{itemize}
In most cases the constraint imposed by a given symmetry on the
renormalisation matrix $\mathds{Z}$ can be easily found out, while in
a few cases (namely heavy quark spin symmetry and $H(3)$ spatial
rotations) an explicit construction of the corresponding $\Phi_k$
matrices is required. We leave the explanation of this procedure to
Appendix~\ref{app:A} and we present here the list of symmetries that
have been used and their effect in constraining the matrix
$\mathds{Z}$.
\vskip 0.3cm
\vskip 0.3cm
\noindent {\bf Flavour exchange symmetry ${\cal S}$}. ${\cal S}$
exchanges the two relativistic flavours $\psi_{1}$ and $\psi_2$. Operators with 
superscript $\pm$ are eigenvectors of $\Phi_{\cal S}$ with eigenvalues
$\pm 1$ respectively. ${\cal S}$ thus prevents the
mixing between the $+$ and $-$ sectors. This reduces the
renormalisation matrix $\dZ$ to a block-diagonal form,
with two $8\times 8$ blocks.
\vskip 0.3cm
\noindent {\bf Parity}. Mixing among operators with opposite parity is
excluded, and the renormalisation matrix $\dZ$ is reduced to a
block-diagonal form, where four $4\times 4$ blocks describe the mixing
of the parity-even and parity-odd operators among themselves.
\vskip 0.3cm
\noindent {\bf Chiral symmetry}. It is used in the same way as in
  ref.~\cite{Donini:1999sf}.
  In the chiral limit, the continuum relativistic quark action 
  is invariant under the finite axial transformation:
\begin{gather}
\psi_k \rightarrow i\gamma_5\psi_k\,;\qquad\bar\psi_k \rightarrow i\bar\psi_k\gamma_5 \ .
\end{gather}
Under this transformation we obtain:
\begin{align}
&  Q^\pm_1 \rightarrow - Q^\pm_1~,      &  \cQ^\pm_1
  \rightarrow - \cQ^\pm_1~, \nonumber\\
&  Q^\pm_2 \rightarrow - Q^\pm_2~,       &  \cQ^\pm_2
  \rightarrow - \cQ^\pm_2~, \nonumber\\
&  Q^\pm_3 \rightarrow\phantom{-} Q^\pm_3~,   & \cQ^\pm_3
  \rightarrow\phantom{-} \cQ^\pm_3~, \nonumber\\
&  Q^\pm_4 \rightarrow\phantom{-} Q^\pm_4~,   & \cQ^\pm_4
  \rightarrow\phantom{-} \cQ^\pm_4~.
\end{align}
From this one sees that, were chirality respected by the regulator,
$Q^\pm_1$ would mix only with $Q^\pm_2$, and $Q^\pm_3$ only with
$Q^\pm_4$ (and similarly in the parity-odd sector). This is not the
case with a Wilson regularisation, for which the structure of chiral
multiplets must be restored by combining operators with different
naive chiralities~\cite{Bochicchio:1985xa}. The restoration of chiral
properties is achieved by introducing the mixing matrices
$\pcd^\pm,\pvd^\pm$. Once the subtracted operators $\tilde Q^\pm =
(\mathds{1}+\pcd^\pm)Q^\pm$ and $\tilde \cQ^\pm =
(\mathds{1}+\pvd^\pm)\cQ^\pm$ with the correct chiral properties have
been constructed, they will mix like in the continuum with
renormalisation matrices $Z^\pm,\cZ^\pm$. We choose the matrices
$Z^\pm,\cZ^\pm,\pcd^\pm,\pvd^\pm$ such that:
\begin{gather}
\left(\begin{array}{c}
	  Q_1^\pm \\
	  Q_2^\pm \\
	  Q_3^\pm \\
	  Q_4^\pm
	\end{array}\right)_{\rm R}
	=
	\left(\begin{array}{cccc}
	  Z_{11}^\pm & Z_{12}^\pm & 0          & 0          \\
	  Z_{21}^\pm & Z_{22}^\pm & 0          & 0          \\
	  0          & 0          & Z_{33}^\pm & Z_{34}^\pm \\
	  0          & 0          & Z_{43}^\pm & Z_{44}^\pm
	\end{array}\right)
	{\left[ \mathds{1} + 
	\left(\begin{array}{cccc}
	  0               & 0                & \pcd_{13}^\pm & \pcd_{14}^\pm \\
	  0               & 0                & \pcd_{23}^\pm & \pcd_{24}^\pm \\
	  \pcd_{31}^\pm & \pcd_{32}^\pm  & 0               & 0 \\
	  \pcd_{41}^\pm & \pcd_{42}^\pm  & 0               & 0
	\end{array} \right)
	\right]}
	\left(\begin{array}{c}
	  Q_1^\pm \\
	  Q_2^\pm \\
	  Q_3^\pm \\
	  Q_4^\pm
	\end{array}\right),
\end{gather}
and
\begin{gather}
\left(\begin{array}{c}
	  \cQ_1^\pm \\
	  \cQ_2^\pm \\
	  \cQ_3^\pm \\
	  \cQ_4^\pm
	\end{array}\right)_{\rm R}
	=
	\left(\begin{array}{cccc}
	  \cZ_{11}^\pm & \cZ_{12}^\pm & 0          & 0          \\
	  \cZ_{21}^\pm & \cZ_{22}^\pm & 0          & 0          \\
	  0          & 0          & \cZ_{33}^\pm & \cZ_{34}^\pm \\
	  0          & 0          & \cZ_{43}^\pm & \cZ_{44}^\pm
	\end{array}\right)
	{\left[ \mathds{1} + 
	\left(\begin{array}{cccc}
	  0               & 0                & \pvd_{13}^\pm & \pvd_{14}^\pm \\
	  0               & 0                & \pvd_{23}^\pm & \pvd_{24}^\pm \\
	  \pvd_{31}^\pm & \pvd_{32}^\pm  & 0               & 0 \\
	  \pvd_{41}^\pm & \pvd_{42}^\pm  & 0               & 0
	\end{array} \right)
	\right]}
	\left(\begin{array}{c}
	  \cQ_1^\pm \\
	  \cQ_2^\pm \\
	  \cQ_3^\pm \\
	  \cQ_4^\pm
	\end{array}\right).
\end{gather} 
This choice is convenient because it is easy to show (for instance by
using Ward identities) that $\pcd^\pm,\ \pvd^\pm$ and
the product $\cZ^\pm(Z^\pm)^{-1}$ all depend only on the bare 
coupling $g_0$, while
$Z^\pm$ and $\cZ^\pm$ alone contain the continuum-like dependence on
the renormalisation scale.
\vskip 0.3cm
\noindent {\bf Heavy quark spin symmetry and $H(3)$ spatial
    rotations}. Further constraints can be obtained from the heavy
    quark spin symmetry and cubic rotations. The procedure is
    slightly involved and we leave its description to
    Appendix~\ref{app:A}. It applies identically to both parity-even
    and parity-odd sectors, and below we provide the expressions for
    the latter --- results for the parity-even sector are obtained by
    simply replacing the symbols $\cQ,\cZ,\pvd$ with $Q,Z,\pcd$.
    After imposing the constraints $\dZ=\Phi_k\dZ\Phi_k^{-1}$ one
    finds that it is possible to rotate~(\ref{basis}) into a new basis
\begin{equation}
\label{diagbasis}
(\cQ'^\pm_1,\cQ'^\pm_2,\cQ'^\pm_3,\cQ'^\pm_4)^T = (
  \cQ^\pm_1,\cQ^\pm_1+4\cQ^\pm_2,\cQ^\pm_3+2\cQ^\pm_4,\cQ^\pm_3-2\cQ^\pm_4)^T, 
\end{equation}
in which the scale-dependent mixing is completely disentangled (even
though some scale-independent mixing remains):
\begin{gather}
\label{diagmixing}
\left(\begin{array}{c}
	  \cQ_1'^\pm \\
	  \cQ_2'^\pm \\
	  \cQ_3'^\pm \\
	  \cQ_4'^\pm
	\end{array}\right)_{\rm R}
	=
	\left(\begin{array}{cccc}
	  \cZ_{1}'^\pm & 0           & 0           & 0          \\
	  0          & \cZ_{2}'^\pm  & 0           & 0          \\
	  0          & 0           & \cZ_{3}'^\pm  & 0 \\
	  0          & 0           & 0           & \cZ_{4}'^\pm
	\end{array}\right)
	{\left[ \mathds{1} + 
	\left(\begin{array}{cccc}
	  0          & 0           & \pvd_{1}'^\pm & 0 \\
	  0          & 0           & 0          & \pvd_{2}'^\pm \\
	  \pvd_{3}'^\pm & 0           & 0          & 0 \\
	  0          & \pvd_{4}'^\pm  & 0          & 0
	\end{array}\right)
	\right]}
	\left(\begin{array}{c}
	  \cQ_1'^\pm \\
	  \cQ_1'^\pm \\
	  \cQ_3'^\pm \\
	  \cQ_3'^\pm
	\end{array}\right) .
\end{gather}
\vskip 0.3cm
\noindent {\bf Time reversal}. Up to this point, the renormalisation mixing
  of the parity-even and parity-odd sectors has the same
  matrix structure. We now consider the effect of a time reversal
  transformation of both the static and relativistic quark fields.
  To that purpose, it is convenient to rewrite (\ref{genop}) in the form
\begin{equation}
\cO^\pm_{\Gamma_1\Gamma_2} =
\frac{1}{2}\left[\left(\bar\Psi P_\lp\Gamma_1\psi_1\right)\left(\bar\Psi
  P_\lm\Gamma_2\psi_2\right) \pm \left(\bar\Psi
  P_\lm\Gamma_1\psi_2\right)\left(\bar\Psi P_\lp\Gamma_2\psi_1\right)\right]\ ,
\end{equation}
with
\begin{equation}
\Psi = \psi_h +
  \psi_{\bar h}\ , \qquad
\bar\Psi = \bar\psi_h + \bar\psi_{\bar h}\ .
\end{equation}
It is then easy to apply the time reversal transformation
\begin{gather}
\begin{split}
 \bar\Psi(x) &\to \bar\Psi(x^\tau)\gamma_5\gamma_0, \\
 \psi_k(x) &\to \gamma_0\gamma_5\psi_k(x^\tau),\quad k=1,2
\end{split}
\end{gather}
where $(x_0,{\bf x})^\tau=(-x_0,{\bf x})$.

The transformation rules for the operators (apart from the $\tau$ reflection 
of the space-time coordinates) are easily found to be
\begin{eqnarray}
  Q_1^\pm \to \pm Q_1^\pm &&   \cQ_1^\pm \to \mp\cQ_1^\pm \ ,\nonumber \\
  Q_2^\pm \to \pm Q_2^\pm &&  \cQ_2^\pm \to \mp\cQ_2^\pm \ , \nonumber\\
  Q_3^\pm \to \pm Q_3^\pm &&  \cQ_3^\pm \to \pm\cQ_3^\pm \ , \nonumber\\
  Q_4^\pm \to \pm Q_4^\pm &\qquad\qquad\qquad\qquad&  \cQ_4^\pm \to \pm\cQ_4^\pm 
\end{eqnarray}
and identical ones for the $(Q',\cQ')$ basis.
It is then clear that no new constraints arise for the
parity-even operators and for $\cZ'^\pm$, while the invariance of 
the scale independent mixing under time reversal immediately yields
\begin{gather}
\pvd'^\pm = 0.
\end{gather}
This proves the purely multiplicative renormalisability of the
operator basis (\ref{diagbasis}). As a note of reference, we would
like to point out that the absence of mixing is already manifest at
the one-loop perturbative level in eqs.~(20)--(23) of
\cite{Gimenez:1998mw}: it is enough to project both sides onto parity
eigenstates and change to the basis in eq.~(\ref{diagbasis}) to find
that mixing in the parity-odd sector is absent at one loop.

Finally, we can rotate back to the standard
basis~(\ref{basis}), obtaining the following form for the
renormalisation matrices:
\begin{align}
  \label{z_pc}
  Z^\pm &= \left(\begin{array}{cccc}
    Z'^\pm_1 & 0 & 0 & 0 \\
    -\frac{1}{4}(Z'^\pm_1-Z'^\pm_2) & Z'^\pm_2 & 0 & 0 \\
    0 & 0 & \frac{1}{2}(Z'^\pm_3+Z'^\pm_4) & Z'^\pm_3-Z'^\pm_4 \\
    0 & 0 & \frac{1}{4}(Z'^\pm_3-Z'^\pm_4) & \frac{1}{2}(Z'^\pm_3+Z'^\pm_4)
    \end{array}\right) \ , \\
  \label{d_pc}
  \pcd^\pm &= \left(\begin{array}{cccc}
    0 & 0 & \pcd'^\pm_1 & 2\pcd'^\pm_1 \\
    0 & 0 & -\frac{1}{4}(\pcd'^\pm_1-\pcd'^\pm_2) & -\frac{1}{2}(\pcd'^\pm_1+\pcd'^\pm_2) \\
    \frac{1}{2}(\pcd'^\pm_3+\pcd'^\pm_4) & 2\pcd'^\pm_4 & 0 & 0 \\
    \frac{1}{4}(\pcd'^\pm_3-\pcd'^\pm_4) & -\pcd'^\pm_4 & 0 & 0
    \end{array}\right) \ ,\\
  \label{z_pv}
  \cZ^\pm &= \left(\begin{array}{cccc}
    \cZ'^\pm_1 & 0 & 0 & 0 \\
    -\frac{1}{4}(\cZ'^\pm_1-\cZ'^\pm_2) & \cZ'^\pm_2 & 0 & 0 \\
    0 & 0 & \frac{1}{2}(\cZ'^\pm_3+\cZ'^\pm_4) & \cZ'^\pm_3-\cZ'^\pm_4 \\
    0 & 0 & \frac{1}{4}(\cZ'^\pm_3-\cZ'^\pm_4) & \frac{1}{2}(\cZ'^\pm_3+\cZ'^\pm_4)
    \end{array}\right) \ ,\\
  \label{d_pv}
    \pvd^\pm &= 0\, .
\end{align}
For later convenience, we denote by $\Lambda$ the matrix responsible
for the change of basis (\ref{diagbasis}), such that $\cQ_i'^\pm =
\Lambda_{ij}\cQ_j^\pm$ and $Q_i'^\pm =
\Lambda_{ij}Q_j^\pm$. Equation~(\ref{z_pc}) reproduces the result
of~\cite{Becirevic:2003hd}.

Now we will pursue a strategy to compute the renormalisation matrices
non perturbatively using Schr\"odinger functional techniques. For parity-even operators, the determination of the subtraction coefficients in eq.~(\ref{d_pc}) can be achieved by implementing suitable axial Ward identities. However, as we will show later on, the use of tmQCD techniques allows to obtain all the $\Delta B=2$ physical amplitudes of interest in the Standard Model from matrix elements of the operators $\cQ_1^+,\cQ_2^+$. Therefore, from now on we will concentrate exclusively on the renormalisation of $\cQ_k^\pm$.

%% file: renorm_cond.tex
\section{Renormalisation conditions in the Schr\"odinger functional
  \label{sec:ren_cond}}

Having constrained their renormalisation patterns by imposing the
symmetries of the theory, we now proceed to specifying suitable
renormalisation conditions on the four-quark operators
(\ref{diagbasis}). To this end we will consider a set of correlation
functions defined in the Schr\"odinger functional (SF) which will
serve to impose the renormalisation conditions at the non-perturbative
level. The SF formalism \cite{Luscher:1992an}, which was initially
developed to produce a precise determination of the running coupling
\cite{alpha:su3,mbar:pap1,DellaMorte:2004bc}, has been extended to
various other phenomenological contexts. These include the study of
quark masses \cite{mbar:pap3,RolfSint_mc,DellaMorte:2005kg} and decay
constants \cite{mbar:pap2,juettner_fDs,deDivitiis:2003wy}, the
computation of moments of structure functions
\cite{Guagnelli:2004ga,Guagnelli:2004ww}, the Kaon
$B$-parameter~\cite{Guagnelli:2005zc,Palombi:2005zd,Dimopoulos:2006dm},
and the static-light axial current~\cite{Kurth:2000ki,Heitger:2003xg}.
The reader is referred to \cite{Luscher:1996sc} for detailed
explanations of the framework and the standard notation, which are not
repeated here.

Our construction of the relevant SF correlators extends the one described in
\cite{Palombi:2005zd} to the static case. We start by introducing the following bilinear
boundary source operators, projected to zero external momentum,
\begin{align}
\label{sources}
\Sigma_{\fla_1\fla_2}[\Gamma] &=
a^6\sum_{\vx,\vy}\bar{\zeta}_{\fla_1}(\vx)\Gamma\zeta_{\fla_2}(\vy) \
, \nonumber \\[1.5ex]
\Sigma'_{\fla_1\fla_2}[\Gamma] &= a^6\sum_{\vx,\vy}\bar{\zeta}'_{\fla_1}(\vx)\Gamma\zeta'_{\fla_2}(\vy) \ ,
\end{align}
where $\Gamma$ is a Dirac matrix, the flavour indices $\fla_{1,2}$ can
assume both relativistic and static values, and the fields $\zeta$ and
$\zeta'$ represent functional derivatives with respect to the
fermionic boundary fields of the SF, see \cite{Luscher:1996sc}. The
choices for $\Gamma$ are limited by the boundary conditions imposed on
quark fields. At the level of boundary quark and antiquark fields they
imply
\begin{gather}
\zeta(\vx)=P_-\zeta(\vx) \, ,~~~~~~~~~~~~\bar\zeta(\vx)=\bar\zeta(\vx)P_+ \, ,
\end{gather}
with $P_\pm=\half(\mathds{1}\pm\gamma_0)$, and similarly for
$\zeta',\bar\zeta'$. Therefore, in order to have a non-vanishing source
field $\Sigma_{\fla_1\fla_2}[\Gamma]$, $\Gamma$ must anticommute with
$\gamma_0$.\footnote{Allowing for a non-vanishing angular momentum would
relax this constraint, but, since it would most likely lead also to
worse signal-to-noise ratios in numerical simulations, we do not
pursue this approach.}

\begin{figure}[!t]
\begin{center}
\epsfig{figure=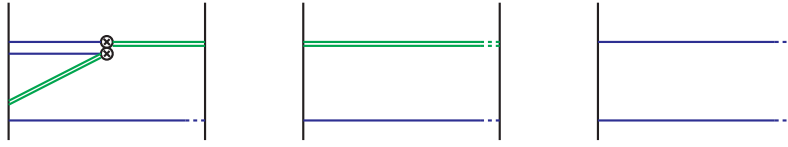, width=10.5 true cm}
\end{center}
\caption{Diagrammatic representation of correlation functions:
  the four-quark correlator ${\cal F}_k^{\pm;(s)}(x_0)$ (left), the 
  boundary-to-boundary static-light correlators $f_{1,hl},k_{1,hl}$ 
  (center) and the boundary-to-boundary light-light correlators
  $f_{1,ll},k_{1,ll}$ (right). Euclidean time goes from left to
  right. The double blob indicates the four-quark operator
  insertion. Single lines represent relativistic quarks and double
  lines denote the static ones. \label{fig_corrfns}} 
\end{figure}

Starting from the bilinears in (\ref{sources}), we have to construct
suitable boundary sources to probe four-quark operators. In order to
define a set of non-zero correlators in the massless theory, which can
then be used to impose renormalisation conditions in the chiral limit,
the probe should be parity-odd. We further require it to be invariant
under the group $H(3)$ of lattice rotations in three
dimensions. This leads us to introduce a third ``spectator'' light
quark and consider, as the simplest possible choice, a generalised
boundary source made of three bilinears,
\begin{align}
\cW[{\Gamma}_1,{\Gamma}_2,{\Gamma}_3] &=
\Sigma_{1h}[{\Gamma}_1]\Sigma_{23}[{\Gamma}_2]\Sigma'_{3\bar h}[{\Gamma}_3] \ ,
\end{align}
two of which are localised at the boundary $x_0=0$, while the third
lies at the other boundary, namely $x_0=T$. Odd parity and rotational
invariance are then assured through an appropriate choice of the Dirac
structures $[\Gamma_1,\Gamma_2,\Gamma_3]$, and a maximal set of
corresponding probes is given by
\begin{align}
\label{source3_1} \pvs^{(1)} &= \cW[\gamma_5,\gamma_5,\gamma_5] \ , \\[2.1ex]
\label{source3_2} \pvs^{(2)} &= \frac{1}{6}\sum_{k,l,m=1}^3\eps_{klm}\cW[\gamma_k,\gamma_l,\gamma_m] \ , \\
\label{source3_3} \pvs^{(3)} &= \frac{1}{3}\sum_{k=1}^3\cW[\gamma_5,\gamma_k,\gamma_k] \ , \\
\label{source3_4} \pvs^{(4)} &= \frac{1}{3}\sum_{k=1}^3\cW[\gamma_k,\gamma_5,\gamma_k] \ , \\
\label{source3_5} \pvs^{(5)} &= \frac{1}{3}\sum_{k=1}^3\cW[\gamma_k,\gamma_k,\gamma_5] \ .
\end{align}
The four-quark operators can now be treated as local insertions in the
bulk of the SF, and their correlators are naturally defined as
\begin{align}
\label{4F_corr}
\cF^{\pm;(s)}_k(x_0) &= L^{-3}\langle \pvo'^\pm_k(x) \pvs^{(s)}
\rangle,\quad s=1,\ldots,5 \ .
\end{align}
A pictorial interpretation of (\ref{4F_corr}) is provided by the left
diagram of Figure~\ref{fig_corrfns}.

It has to be observed that, due to the symmetries of the static
approximation for the heavy quarks, not all of the above correlation
functions are independent. By using the explicit spin structure of the static
propagator, some straightforward though tedious algebra leads to the
constraints
\begin{alignat}{2}
\label{id_scheme}
\cF_1^{\pm;(4)} &= -\cF_1^{\pm;(1)} \ , \qquad
\cF_1^{\pm;(3)} &= -\cF_1^{\pm;(2)} \ , \nonumber \\
\cF_2^{\pm;(4)} &= \frac{1}{3}\cF_2^{\pm;(1)} \ , \qquad
\cF_1^{\pm;(5)} &= -\cF_1^{\pm;(2)} \ , \nonumber \\
\cF_3^{\pm;(4)} &= -\cF_3^{\pm;(1)} \ , \qquad
\cF_3^{\pm;(3)} &= -\cF_3^{\pm;(2)} \ , \nonumber \\
\cF_4^{\pm;(4)} &= \frac{1}{3}\cF_4^{\pm;(1)} \ , \qquad
\cF_3^{\pm;(5)} &= -\cF_3^{\pm;(2)} \ .
\end{alignat}
These relations show that only 24 out of the 40 correlation functions defined
in eq.~(\ref{4F_corr}) are independent. We stress that this result is exact; 
as a cross-check, later on we will find these identities to be explicitly
verified at one loop in perturbation theory.

In order to isolate from eq.~(\ref{4F_corr}) the ultraviolet
divergences that are due to the bulk operator and absorb them into a
renormalisation factor, one has to address the renormalisation of the
boundary fields. The ultraviolet divergences of the latter are
cancelled by defining suitable ratios of correlators for which the
renormalisation factors of the boundary fields drop out. To this end
we introduce a set of boundary-to-boundary light-light and
static-light correlators,
\begin{align}
f_1^{hl} &= -\frac{1}{2L^6}\langle\cO'_{1\bar h}[\gamma_5]\cO_{h1}[\gamma_5]\rangle \ ,\\[1.5ex]
f_1^{ll} &=
-\frac{1}{2L^6}\langle\cO'_{12}[\gamma_5]\cO_{21}[\gamma_5]\rangle \ , \\[1.5ex]
k_1^{ll} &= -\frac{1}{6L^6}\sum_{k=1}^3\langle\cO'_{12}[\gamma_k]\cO_{21}[\gamma_k]\rangle\ , 
\end{align}
whose valence structure is represented as well in the middle and right
diagrams of Figure~\ref{fig_corrfns}.~\footnote{The static-light
  counterpart of $k_1^{ll}$ is not considered as, due to the symmetries
  of the static limit, it is identical to $f_1^{hl}$.} 
A suitable combination of such
correlators must comprise the same number of static and light boundary
fields as (\ref{4F_corr}). The simplest examples are given by the
ratios
\begin{equation}
 \frac{{\cal F}_k^{\pm;(s)}(x_0)}{f_1^{hl}[f_1^{ll}]^{1/2}},\qquad
 \frac{{\cal F}_k^{\pm;(s)}(x_0)}{f_1^{hl}[k_1^{ll}]^{1/2}},
\end{equation}
involving either the correlator $f_1^{ll}$ or $k_1^{ll}$. However, it
is easy to generalise these conditions by introducing an arbitrary
real parameter $\alpha$. Hence, we consider the following ratios
of correlation functions:
\begin{align}
\label{h_ratio}
h_{k;\alpha}^{\pm;(s)}(x_0) &= \frac{{\cal
    F}_k^{\pm;(s)}(x_0)}{f_1^{hl}[f_1^{ll}]^{1/2-\alpha}[k_1^{ll}]^\alpha}\ .
\end{align}
Although there is no real {\em a priori}
restriction on the value of $\alpha$, it is clear that ``natural''
values should lie in the interval $[0,\half]$. This freedom,
together with the choice of the boundary source and the
$\theta$-angle of the SF, can be used in a later stage to tune the 
optimal renormalisation schemes, with the aim of having small NLO
coefficients in the corresponding anomalous dimensions. This is
important in order to control the systematics of the perturbative
matching to continuum schemes at high energy scales. 

For the moment we observe that the
ratios (\ref{h_ratio}) are free of boundary divergences, and 
consequently we impose the renormalisation condition,
\begin{align}
\label{ren_cond}
\cZ'^{\pm;(s)}_{k;\alpha} h_{k;\alpha}^{\pm;(s)}(T/2) &=
h_{k;\alpha}^{\pm;(s)}(T/2)|_{g_0=0}\ ,
\end{align}
where all the correlation functions are computed in the chiral limit.
This fixes non-perturbatively the renormalisation constant $\cZ'^{\pm;(s)}_{k;\alpha}$ at the
scale $\mu = 1/L$. As usual, the $\cZ'$ factors depend upon every
calculational detail with the only exception of the leading log, which
is universal. In order to operatively define a renormalisation scheme,
a complete specification of the parameters that concur to quantify
(\ref{h_ratio}) is required. We briefly summarise them:
\begin{itemize}
\item the possible presence of a Sheikholeslami-Wohlert (SW) term in
  the lattice action for the light quarks; 
\item the choice of the gauge parallel transporter in the covariant
  derivatives, eq.\,(\ref{eq_latt_der}), in the static action;
\item the Dirac structure of the boundary source;
\item the value of the angle $\theta$ entering the spatial boundary
  conditions of the SF;
\item the value of the parameter $\alpha$ in \eq{h_ratio};
\item the ratio $T/L$ between the time and the spatial extension of
 the SF.
\end{itemize}
The last four conditions fix the renormalisation scheme, while the first two
only introduce a regularisation dependence in the renormalisation constants.
We stress at this point that the running of the operators is a continuum
property, i.e. it is independent of the discretisation chosen. The latter
only affects the way in which the continuum limit is approached, and -- in the
case of the action for static quark fields -- the signal-to-noise ratio in
the simulations.

In this paper we consider the Wilson action (with and without SW term) for the
light quarks, and the Eichten-Hill action for the static ones. As for the
parameters that fix the renormalisation scheme, we will consider three values
of $\theta$ (namely $\theta=0.0,~0.5,~1.0$) and two values of $\alpha$ (namely
$\alpha=0.0,~0.5$). Furthermore, we fix $T=L$. Taken together with the possible
independent choices of boundary sources, this leaves us with 12 different
renormalisation schemes for the operators $\cQ'^\pm_1$ and $\cQ'^\pm_3$,
and 24 schemes for $\cQ'^\pm_2$ and $\cQ'^\pm_4$. The number of independent
renormalisation conditions is twice this figure, as we consider two different
actions.

%% file: perturb.tex
\section{A perturbative study \label{sec:perturb}}

We now proceed to studying the renormalisation of the operators 
$\cQ_1^+,\ldots,\cQ_4^+$ at one-loop in perturbation theory. These
are the operators that will enter $\Delta B=2$ effective Hamiltonians
in the static limit. The purpose of this study is threefold.
First, it provides an explicit check of the expected mixing pattern.
Second, it will allow us to compute the NLO anomalous dimension of
the operators in the SF schemes defined above; this is done via a standard
one-loop matching procedure to continuum schemes where the NLO anomalous 
dimension is already known. Finally, we can work out lattice
artefacts of the step scaling functions in one-loop perturbation
theory. They can then be compared to the ones in the relativistic case
discussed in \cite{Palombi:2005zd}, in order to obtain information
about the expected size of discretisation effects for quantities
involving static fields. In principle, the discretisation effects
computed in perturbation theory could subsequently be subtracted by
hand from the non-perturbative Monte Carlo data, with the aim of
exerting a better control of their continuum extrapolation, as pursued
in \cite{deDivitiis:1994yz,Guagnelli:2004za,Guagnelli:2005zc}.

\subsection{Scheme dependence of NLO anomalous dimensions}

In order to proceed, some notation has to be fixed. The scale
dependence of operator insertions in renormalised correlation
functions is described by the RG equation
\begin{equation}
 \left[\delta_{ij}\left(\mu\frac{\partial}{\partial\mu} +
  \beta\frac{\partial}{\partial g} +
  \beta_\lambda\lambda\frac{\partial}{\partial \lambda} +
  \tau m\frac{\partial}{\partial m}\right) -
  \gamma_{ij}^\lp\right](\cQ^+_j)_{\rm R} = 0\ ,
\end{equation}
where $\beta$ is the $\beta$-function for the coupling, $\tau$ is the
anomalous dimension of the quark mass, and $\gamma^+$ is the operator
anomalous dimension matrix. We have also included a term which takes 
into account the dependence
on the gauge parameter $\lambda$ in covariant gauges, characterised by
the RG function $\beta_\lambda$, defined as
\begin{gather}
\mu\frac{\partial\lambda}{\partial\mu} = \lambda\beta_\lambda \, .
\end{gather}
This term is absent in schemes like \MSbar (irrespective of the
regularisation prescription) or the SF schemes introduced in
section~\ref{sec:ren_cond}, but is present e.g. in
regularisation-independent (RI) schemes, which will be considered
later on. If we choose to work with the operator basis in
eq.~(\ref{diagbasis}), the matrix structure of the RG-equation
simplifies, and the evolution of the various operators is determined
by a set of scalar anomalous dimensions.

In what follows, we focus on mass-independent renormalisation schemes,
for which the RG functions depend only upon the coupling. We take the
following form for their perturbative expansions:
\begin{alignat}{3}
\beta(g) & = & -g^3\sum_{k=1}^\infty b_kg^{2k}\ , \qquad
\beta_\lambda(g) & = -g^2\sum_{k=0}^\infty b_k^\lambda g^{2k} \ ,\nonumber \\
\tau(g) & = & -g^2\sum_{k=0}^\infty d_kg^{2k}\ ,  \qquad
\gamma_{ij}^+(g) & = -g^2\sum_{k=0}^\infty \gamma^{\lp;(k)}_{ij}g^{2k}\ ,
\end{alignat}
with universal coefficients
\begin{align}
b_0 & = \frac{1}{(4\pi)^2}\left\{\frac{11}{3}N - \frac{2}{3}N_{\rm
  f}\right\}\ , \qquad d_0 = \frac{1}{(4\pi)^2}\left\{3\frac{N^2-1}{N}\right\} \ ,
  \nonumber \\[1.5ex]
b_0^\lambda & = \frac{1}{(4\pi)^2}\left\{N\left(\lambda - \frac{13}{3}\right) + \frac{4}{3}N_{\rm f}\right\}\ , \nonumber
\\[1.5ex]
b_1 & = \frac{1}{(4\pi)^4}\left\{\frac{34}{3}N^2 -
  \left(\frac{13}{3}N - N^{-1}\right)N_{\rm f}\right\}\ .
\end{align}
The LO coefficient of the anomalous dimension matrix
$\gamma^{+;(0)}$ is universal as well, and has been
calculated in \cite{Flynn:1990qz}
for the first operator of the basis and
\cite{Gimenez:1998mw} for the rest. The non-vanishing elements
of the anomalous dimension matrix read
\begin{align}
\gamma_{11}^{+;(0)} &= -\frac{1}{(4\pi)^2}\left(3N-\frac{3}{N}\right) \ ,\\
\gamma_{21}^{+;(0)} &=  \frac{1}{(4\pi)^2}\left(1+\frac{1}{N}\right) \ , \\
\gamma_{22}^{+;(0)} &= -\frac{1}{(4\pi)^2}\left(3N-4-\frac{7}{N}\right) \ , \\
\gamma_{33}^{+;(0)} &= -\frac{1}{(4\pi)^2}\left(3N-\frac{6}{N}\right) \ , \\
\gamma_{34}^{+;(0)} &= -\frac{6}{(4\pi)^2} \ ,\\
\gamma_{43}^{+;(0)} &= -\frac{3/2}{(4\pi)^2} \ ,\\
\gamma_{44}^{+;(0)} &= -\frac{1}{(4\pi)^2}\left(3N-\frac{6}{N}\right) \ .
\end{align}
A covariant rotation of this matrix to the
diagonal operator basis (\ref{diagbasis}), i.e.
$\gamma'^{+;(0)} = \Lambda \gamma^{+;(0)} \Lambda^{-1}$, gives the LO coefficients of
the multiplicatively renormalisable operators, namely
\begin{alignat}{3}
\label{gamma0}
\gamma_1'^{+;(0)} & =  - \frac{1}{(4\pi)^2}\left(3N - \frac{3}{N}\right)\ , \quad\qquad
& \gamma_2'^{+;(0)} & = -\frac{1}{(4\pi)^2}\left(3N-4-\frac{7}{N}\right)\ , \nonumber \\[1.7ex]
\gamma_3'^{+;(0)} & =  - \frac{1}{(4\pi)^2}\left(3N+3-\frac{6}{N}\right) \ , \quad\qquad
& \gamma_4'^{+;(0)} & = -\frac{1}{(4\pi)^2}\left(3N - 3 -
\frac{6}{N}\right) \ .
\end{alignat}

By contrast, the NLO coefficient is scheme-dependent. The perturbative
matching procedure that allows to express its value in the SF scheme
in terms of the value in a reference scheme has been derived in
\cite{Sint:1998iq} for the case of multiplicatively renormalisable
operators. The formalism can be trivially extended to situations where
mixing occurs and a gauge covariant reference scheme is assumed. The
renormalised operators and coupling constant are first related in the
two schemes through a finite renormalisation,
\begin{align}
\label{schemechange}
  g^2_{\rm SF} & = \cX_g(g_{\rm ref})g^2_{{\rm ref}}\ , \nonumber \\[1.8ex]
  (\cQ^+_{i,{\rm SF}})_{\rm R} & = \cX^+_{ij}(g_{\rm ref})(\cQ^+_{j,\rm ref})_{\rm R} \ .
\end{align}
The matching coefficients $\cX$ are then expanded in powers of
the coupling constant,
\begin{equation}
\cX(g) = 1 + \sum_{k=1}^\infty g^{2k}\cX^{(k)}\ ,
\end{equation}
and the requirement of formal invariance of the RG-equation under a
change of renormalisation scheme leads to the two-loop matching
relation
\begin{equation}
\label{twoloopconn}
\gamma_{\rm SF}^{+;(1)} =
\gamma_{\rm ref}^{+;(1)} + [\cX^{+;(1)},\gamma^{+;(0)}] +
2b_0\cX^{+;(1)}  +  b_0^\lambda\lambda\frac{\partial}
  {\partial \lambda}\cX^{+;(1)}  - \gamma^{+;(0)}\cX_{g}^{(1)}\ ,
\end{equation}
where the symbol $[\cdot,\cdot]$, which is absent in the case of
multiplicative renormalisation, represents the ordinary matrix
commutator. It should be stressed that the choice of the reference
scheme is irrelevant. In fact, a good consistency check on the result
for $\gamma_{\rm SF}^{+;(1)}$ is provided by computing the RHS of
(\ref{twoloopconn}) for several different reference schemes.
\vskip 0.3cm Finally, we point out that the lattice is currently the
only known regularisation of the SF, for which perturbative
calculations of fermionic observables can be operatively
performed.\footnote{A recent proposal to perform the matching directly in
dimensional regularisation has been presented in \cite{Obeso:2005mc}.}
If the reference scheme is defined in the continuum, the operator
matching must take into account both a change of regularisation and a
change of subtraction prescription. Accordingly, $\cX_\cO^{(1)}$ must
be computed as the difference of two matching coefficients to an
intermediate scheme, namely
\begin{equation}
\label{intermediatematch}
\cX_{{\rm SF},\rm ref}^{+;(1)} =
\cX_{{\rm SF},\rm lat}^{+;(1)} -
\cX_{{\rm ref},{\rm lat}}^{+;(1)} \ .
\end{equation}
The ``lat'' scheme is by definition the minimal subtraction lattice
scheme, where the renormalisation constants are polynomials in
$\ln(a\mu)$ without finite parts. Consequently, $\cX_{\rm
SF,lat}^{+;(1)}$, which provides the matching between SF and ``lat'',
can be obtained from a one-loop calculation of the renormalisation
constant in the SF scheme with a lattice regularisation. The matching
coefficient $\cX_{\rm ref,lat}^{+;(1)}$ between the reference scheme
and the lattice can be instead retrieved from the literature for some
choice of the reference scheme, such as \MSbar or RI.

\subsection{Perturbative expansion of SF correlation functions
  \label{sec:pert_exp}} 

We now describe the one-loop calculation of the SF renormalisation
constants introduced in section~\ref{sec:ren_cond}. The perturbative
procedure is fairly conventional, and we include it just for
completeness. We start by expanding all the correlation functions
previously introduced in powers of the bare coupling,
\begin{align}
\label{X_exp}
X = X^{(0)} + g_0^2\left[X^{(1)} + m_c^{(1)}\frac{\partial
    X^{(0)}}{\partial m_0}\right] + O\left(g_0^4\right),
\end{align}
where $X$ is one of $\cF^{+;(s)}_k$, $f_1^{h l}$,
$f_1^{ll}$, $k_1^{ll}$, or a linear combination thereof. The
derivative term in square brackets is required in order to set the
correlation function $X$ to zero renormalised quark mass, when each
contribution to the RHS is calculated at zero bare quark mass, as it
will be assumed. As for the numerical value of $m_c^{(1)}$, we use the
numbers provided by \cite{Panagopoulos:2001fn}, i.e.
\begin{align}
  am_c^{(1)} = \begin{cases} -0.20255651209\,\Cf &\text{($\csw=1$),}
    \\
    -0.32571411742\,\Cf &\text{($\csw=0$),}
  \end{cases}
  \qquad \Cf=\dfrac{N^2-1}{2N},
  \label{eq:mc1}
\end{align}
The SF renormalisation constants, defined in (\ref{ren_cond}), admit an analogous expansion,
\begin{align}
\label{Z_exp}
\cZ'^{+;(s)}_{k;\alpha}(g_0,a/L) = 1 + g_0^2\cZ'^{+;(s,1)}_{k;\alpha}(L/a) + O\left(g_0^4\right).
\end{align}
The explicit expression of the one-loop order coefficient
$\cZ'^{+;(s,1)}_{k;\alpha}$ in terms of the perturbative expansion of
the four-quark and the boundary-to-boundary correlators can be
obtained by inserting (\ref{X_exp}) and (\ref{Z_exp}) into the
renormalisation condition (\ref{ren_cond}). One then obtains
\begin{align}
\label{pert_ren}
\cZ'^{+;(s,1)}_{k;\alpha}(L/a) =
& -\left\{\frac{\cF^{+;(s,1)}_k}{\cF^{+;(s,0)}_k} +
\frac{\cF_{kb}^{+;(s,1)}}{\cF_{k}^{+;(s,0)}} +
\frac{m_c^{(1)}}{\cF_k^{+;(s,0)}}\frac{\partial \cF_k^{+;(s,0)}}{\partial m_0} \right\}_{x_0=T/2}
\nonumber\\[1.5ex]
& +\left\{\frac{f_1^{hl(1)}}{f_1^{hl(0)}} +
\frac{f_{1b}^{hl(1)}}{f_1^{hl(0)}}
+ \frac{m_c^{(1)}}{f_1^{hl(0)}}\frac{\partial f_1^{hl(0)}}{\partial m_0}\right\}
\nonumber \\[1.5ex]
& +\left(\frac{1}{2}-\alpha\right)\left\{\frac{f_1^{ll(1)}}{f_1^{ll(0)}} +
\frac{f_{1b}^{ll(1)}}{f_1^{ll(0)}} + \frac{m_c^{(1)}}{f_1^{ll(0)}}
\frac{\partial f_1^{ll(0)}}{\partial m_0} \right\}
\nonumber \\[1.5ex]
& +\alpha\left\{\frac{k_1^{ll(1)}}{k_1^{ll(0)}} +
\frac{k_{1b}^{ll(1)}}{k_1^{ll(0)}} + \frac{m_c^{(1)}}{k_1^{ll(0)}}
\frac{\partial k_1^{ll(0)}}{\partial m_0}\right\}.
\end{align}
Contributions containing the subscript ``$b$'' arise from the boundary
terms that are required in addition to the SW term in order to achieve
full $\Oa$-improvement of the action in the SF \cite{Luscher:1996sc}.
Obviously, these contributions are not present when the unimproved
Wilson action is chosen for the light quarks. From now on we will set
them to zero also when the action is $\Oa$ improved, as they will not
affect the continuum limit extrapolations involved in the computation
of NLO anomalous dimension, and their contribution to cutoff effects
is negligible.

\begin{figure}[t]
\begin{center}
\epsfig{figure=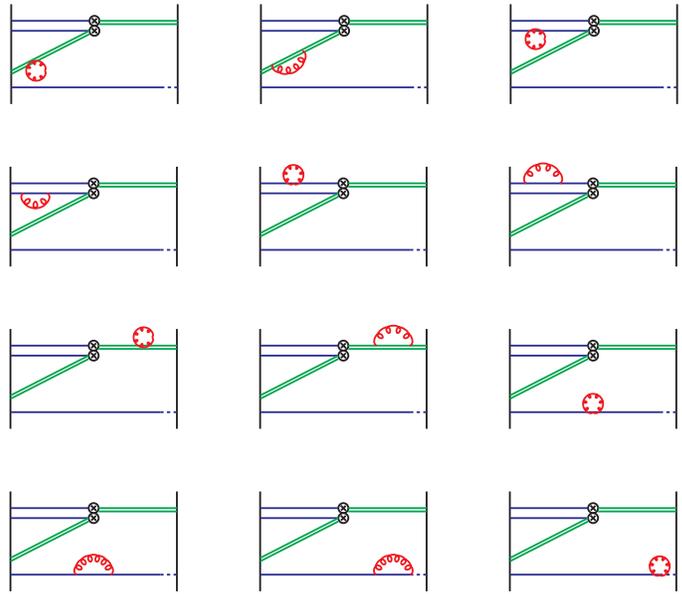, width=8.9 true cm}
\end{center}
\caption{Feynman diagrams of the self-energy type.\label{fig_feyn1}}
\vskip 0.8cm
\end{figure}

\begin{figure}[!h]
\begin{center}
\epsfig{figure=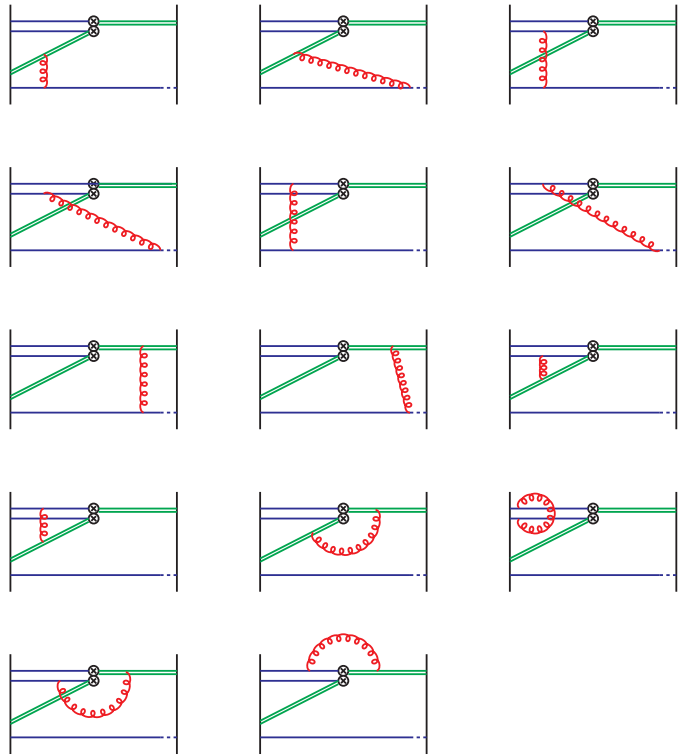, width=8.9 true cm}
\end{center}
\caption{Feynman diagrams of the vertex type.\label{fig_feyn2}}
\end{figure}

The evaluation of the RHS of (\ref{pert_ren}) requires the calculation
of the Feynman diagrams depicted in Figures \ref{fig_feyn1} and
\ref{fig_feyn2}. The one-loop expansion of the boundary-to-boundary
correlators $f_1^{ll}$ and $k_1^{ll}$ is known from
\cite{Sint:1997dj}, while $f_1^{hl}$ has been studied perturbatively
in \cite{Kurth:2000ki}. Accordingly, the only new diagrams which need
to be calculated are the ones that contribute to the one-loop order
coefficient of the four-quark correlators. Two groups of diagrams can
be identified: the {\it self-energies} correct the valence fermion
propagators through a gluon emission with subsequent absorption by the
same leg, and the {\it vertex} diagrams correct the operator
insertions through the exchange of a gluon between two legs. Each of
them can be expressed as a loop sum of a Dirac trace in
time-momentum representation, where the spatial coordinates are
Fourier transformed. These sums have been performed numerically in
double precision arithmetics using a C++ code, for all the even
lattice sizes ranging from $L/a = 6$ to $L/a = 48$. The results have
been checked by an independent Fortran 90 program, also in double
precision arithmetics. The behaviour of the renormalisation constants
thus obtained, as functions of the lattice size $L/a$, is expected to
conform to the standard asymptotic expansion
\begin{align}
\cZ'^{+;(s,1)}_{k;\alpha}(L/a)
  = \sum_{\nu=0}^\infty\left(\dfrac{a}{L}\right)^\nu
  \left\{r_{k,\nu}^+ + s_{k,\nu}^+\ln(L/a)\right\}\ ,
\label{eq:r0}
\end{align}
which can be used in order to extract the universal LO anomalous
dimensions and the finite constants peculiar to the schemes, that is
to say, the coefficients $s^+_{0}$ and $r^+_{0}$, respectively. The
latter represents the matching coefficient introduced in
(\ref{intermediatematch}) in the diagonal basis, namely
\begin{equation}
\label{matchSF}
\cX'^{(1)}_{k;{\rm SF},\rm lat} = r_{k,0}^+ \, .
\end{equation}
An efficient numerical technique to isolate these coefficients, based
on a blocking procedure of the function at neighbour lattice sizes,
has been introduced in \cite{Luscher:1985wf}. Details about its
application to the case at hand are provided to Appendix
\ref{app:C}. Numerical values of the coefficients $r_{k,0}^+$ for the
various schemes introduced in section~\ref{sec:ren_cond} are reported
in Tables~\ref{tab:constants1}~--~\ref{tab:constants6}.

\subsection{Matching to continuum schemes and consistency checks}

The NLO anomalous dimension matrix of the operators (\ref{basis}) in
continuum schemes can be found in \cite{Gimenez:1998mw}, together with
the one-loop matching relations to the minimal subtraction lattice
scheme. The regularisations employed in \cite{Gimenez:1998mw} are DRED
and NDR, and two possible subtraction prescriptions are considered,
namely \MSbar and RI. An attractive feature of the latter is the
independence of the corresponding anomalous dimension from the choice
of evanescent operators (EO), which complicate the mixing pattern
in $d=4-2\epsilon$ dimensions. As a consequence, it is trivial to
perform a rotation of the anomalous dimension matrix in the RI scheme
to a different basis of the physical operators, such as
(\ref{diagbasis}), without the need to address subtleties related to
the definition of evanescent contributions. The choice of RI as a
reference scheme is therefore convenient in order to make use of the
two-loop matching relation (\ref{twoloopconn}) in the diagonal basis
(\ref{diagbasis}).

Results reported in \cite{Gimenez:1998mw} refer to a perturbative
expansion in powers of the $\overline{\rm MS}$-coupling. We therefore
need the matching coefficient in eq.~(\ref{schemechange}), which
relates $g_{\rm SF}$ to $g_{\overline{\rm MS}}$, to one-loop
order. This has been calculated in \cite{Sint:1995ch} and is given by
\begin{align}
\label{coupmatch}
\cX_g^{(1)} & = 2b_0\ln(\mu L) - \frac{1}{4\pi}\left(c_{1,0} +
c_{1,1}N_{\rm f}\right)\ , \nonumber \\[1.5ex]
c_{1,0} & = 1.25563(4)\ , \quad c_{1,1} = 0.039863(2)\ .
\end{align}
The NLO anomalous dimension of the operator basis (\ref{diagbasis}) in
the Feynman gauge ($\lambda = 1$) and NDR
regularisation\footnote{Although the four-quark operators are
renormalised according to the RI scheme, which is independent of the
regularisation prescription, the strong interaction Lagrangian is
renormalised in \MSbar. This introduces a spurious dependence of the
NLO anomalous dimension upon the choice of the regulator.}, obtained
from the covariant rotation $\gamma'^{(1)} =
\Lambda\gamma^{(1)}\Lambda^{-1}$, is a diagonal matrix whose non-zero
coefficients read
\begin{align}
\label{gamma1}
\gamma_{1;{\rm RI}}'^{+;(1)} = \frac{1}{(4\pi)^4}\bigg[&
-\frac{131+8\pi^2}{12}\NC^2 + 6\NC - \frac{1-2\pi^2}{3} +
\frac{30+4\pi^2}{3\NC} - \frac{57+16\pi^2}{12\NC^2}  \nonumber \\
&+ \NF\left(\frac{5}{3}\NC - \frac{5}{3\NC}\right)
\bigg] \ , \nonumber \\[1.7ex]
\gamma_{2;{\rm RI}}'^{+;(1)} = \frac{1}{(4\pi)^4}\bigg[&
-\frac{131+8\pi^2}{12}\NC^2 + \frac{214}{9}\NC + \frac{301+6\pi^2}{9}
+ \frac{18-4\pi^2}{3\NC} + \frac{87-16\pi^2}{12\NC^2}  \nonumber \\
&+ \NF\left(\frac{5}{3}\NC - \frac{40}{9} - \frac{55}{9\NC}\right)
\bigg] \ , \nonumber \\[1.7ex]
\gamma_{3;{\rm RI}}'^{+;(1)} = \frac{1}{(4\pi)^4}\bigg[&
-\frac{131+8\pi^2}{12}\NC^2 - \frac{83}{6}\NC + \frac{309+8\pi^2}{12}
- \frac{24-4\pi^2}{3\NC} + \frac{21-4\pi^2}{3\NC^2}  \nonumber \\
&+ \NF\left(\frac{5}{3}\NC + \frac{10}{3} - \frac{5}{\NC}\right)
\bigg] \ , \nonumber \\[1.7ex]
\gamma_{4;{\rm RI}}'^{+;(1)} = \frac{1}{(4\pi)^4}\bigg[&
-\frac{131+8\pi^2}{12}\NC^2 + \frac{71}{6}\NC + \frac{309+8\pi^2}{12}
+ \frac{42-4\pi^2}{3\NC} + \frac{33-4\pi^2}{3\NC^2}  \nonumber \\
&+ \NF\left(\frac{5}{3}\NC - \frac{10}{3} - \frac{5}{\NC}\right)
\bigg] \ .
\end{align}
The same rotation can be applied to the one-loop operator matching
matrix $\cX'^{(1)}_{{\rm RI, lat}}$. In this case the analytic
dependence upon the gauge parameter $\lambda$ is needed in order to
account for the derivative term included in the two-loop matching
relation (\ref{twoloopconn}). With $N=3$, one has
\begin{align}
  \label{matchRI}
  \cX_{1;{\rm RI,lat(wilson)}}'^{(1)} & =
  \frac{1}{(4\pi)^2}\left[\frac{10}{3} -\frac{8}{3}\lambda +
    (D_{LL}-D_{RR})\right] \ , \nonumber \\[1.5ex]
  \cX_{2;{\rm RI,lat(wilson)}}'^{(1)} & =
  \frac{1}{(4\pi)^2}\left[\frac{10}{3} -\frac{8}{3}\lambda +
    D_{LL}^S\right] \ , \nonumber \\[1.5ex]
  \cX_{3;{\rm RI,lat(wilson)}}'^{(1)} & =
  \frac{1}{(4\pi)^2}\left[\frac{10}{3} -\frac{8}{3}\lambda +
    \frac{1}{4}\left( 2D_{LR}  +
    2D^S_{LR} -4\bar D_{RL} -\bar D^S_{RL}\right)\right]\ , \nonumber
  \\[1.5ex]
  \cX_{4;{\rm RI,lat(wilson)}}'^{(1)} & =
  \frac{1}{(4\pi)^2}\left[\frac{10}{3} - \frac{8}{3}\lambda +
    \frac{1}{4}\left( 2D_{LR} +
    2D^S_{LR} + 4\bar D_{RL} + \bar D^S_{RL} \right)\right]
\end{align}
for light quarks regularised with the pure Wilson action. If the $\Oa$
improved action is used instead, one has to add to them the matching
factors between the two actions, viz.
\begin{alignat}{3}
\cX_{1;{\rm lat(sw),lat(wilson)}}'^{(1)} &=
\frac{1}{(4\pi)^2}\left[-\frac{4}{3}f^I -\frac{1}{3}v^I -\frac{4}{3}w^I \right ] \ &= 0.038033(2) \ , \nonumber \\
\cX_{2;{\rm lat(sw),lat(wilson)}}'^{(1)}  &=
\frac{1}{(4\pi)^2}\left[-\frac{4}{3}f^I - \frac{2}{9}v^I \right] \ &= 0.040240(2) \ , \nonumber \\
\cX_{3;{\rm lat(sw),lat(wilson)}}'^{(1)} &=
\frac{1}{(4\pi)^2}\left[-\frac{4}{3}f^I + \frac{2}{3}w^I \right] \ &= 0.034253(2) \ , \nonumber \\
\cX_{4;{\rm lat(sw),lat(wilson)}}'^{(1)} &=
\frac{1}{(4\pi)^2}\left[-\frac{4}{3}f^I + \frac{4}{3}w^I \right] \ &= 0.037720(2)
 \ , \label{matchingsw}
\end{alignat}
where the lattice integrals $f^I$, $v^I$ and $w^I$ are discussed in
Appendix~\ref{app:B}. Numerical values of the $D$-coefficients,
expressed in \cite{Gimenez:1998mw} as linear combinations of a basic
set of lattice integrals, are reported in Table \ref{tab:infconst} of
Appendix~\ref{app:B}, where a new computational method to improve
their numerical accuracy is also described. The factors in
eq.~(\ref{matchingsw}) are obtained from the coefficients denoted
$D^I$ in \cite{Gimenez:1998mw}, after subtracting the contributions
coming from the $\Oa$ improvement of the four-fermion operators.
\vskip 0.3cm 
All the ingredients needed to evaluate the RHS of \eq{twoloopconn}
have now been specified. The absence of operator mixing in the
diagonal basis (\ref{diagbasis}) implies that the commutator term in
eq.~(\ref{twoloopconn}) is identically zero. NLO anomalous dimensions
in the previously introduced SF schemes follow from a straightforward
use of eqs.~(\ref{gamma0}), (\ref{intermediatematch}) and
(\ref{matchSF})--(\ref{matchRI}). We have collected the ratios of
$\gamma_{k;\rm SF}'^{+;(1)}$ to the corresponding LO coefficients
$\gamma_k'^{+;(0)}$ in Tables \ref{tab:NLO1} -- \ref{tab:NLO6}. In the
matching we have employed the values of $r_{k,0}^+$ obtained with the
pure Wilson action for light quarks, as they tend to display a better
behaved continuum extrapolation, after the $\Oa$ contributions have
been removed through blocking.

In order to check our results, we have also derived the SF NLO
anomalous dimensions using \MSbar as a reference scheme. The matching
procedure, rather delicate in this case, must take into account the
r\^ole played by the EO in fixing the finite contributions to the NLO
anomalous dimension matrix $\gamma^{\lp;(1)}_{\overline{\rm MS}}$. A
naive rotation of the latter is potentially hazardous without
reconsidering the choice of the EO. An alternative approach is to work
within the original basis (\ref{basis}), to which the results in
\cite{Gimenez:1998mw} refer, and then rotate the one-loop matching
coefficients from the SF to the lattice scheme according to the
inverse rotation $\cX_{\rm SF,lat}^{(1)} = \Lambda^{-1}\cX_{\rm
SF,lat}'^{(1)}\Lambda$. This is certainly possible, as the computation
of such coefficients is performed on the lattice in $d=4$ dimensions,
where no EO contributes. Of course, the commutator term in
eq.~(\ref{twoloopconn}) must be included in this case, whilst the
gauge term proportional to $b_0^\lambda$ is not present. Once the NLO
anomalous dimension matrix has been obtained in the SF scheme, a
straight rotation back to the diagonal basis (\ref{diagbasis}) yields
the scalar coefficients $\gamma_{k;\rm SF}'^{+;(1)}$. This procedure
has been applied using either DRED or NDR regularisations. In
both cases we obtain the same results as with RI in the diagonal
basis.

We have also verified that the difference between the finite parts of
the SF renormalisation constants with improved and unimproved Wilson
light quarks coincides with the values in eq.~(\ref{matchingsw}). The
numerical values of these finite matching constants are indeed in
perfect agreement with the analogous SF quantities.

We finally concentrate on the numerical values of the NLO anomalous
dimension coefficients in the SF. A comparison with the case studied
in \cite{Palombi:2005zd}, where the four-fermion operators contain
only relativistic quark fields, shows that in the present case the
variation of the anomalous dimension due to different choices of the
SF boundary sources in the renormalisation condition is much less
pronounced. Also, the non-perturbative identities in
eq.~(\ref{id_scheme}) are verified explicitly by the one-loop
results. The dependence on the value of the parameter $\alpha$ is very
small, too. Finally, at $\theta=0.5$, which is commonly employed in
non-perturbative studies of SF renormalisation, the values obtained
for the ratios $\gamma_{k;\rm SF}'^{+;(1)}/\gamma_k'^{+;(0)}$ are
relatively small, pointing towards a good convergence of the
perturbative series, save for ${\cal{Q}}_2^\lp$, where they are close
to $-0.5$. The question whether this is a relevant source of
uncertainty in the NLO matching of renormalised
matrix elements to continuum schemes is left for future studies.

\subsection{One-loop order cutoff effects in step-scaling functions}

\begin{figure}[!t]
  \begin{center}
    \begin{minipage}{.45\linewidth}
      \psfrag{xlabel}[c][bc][1][0]{$T/a$}
      \psfrag{ylabel}[c][c][1][0]{$\log_{10}r$}
      \begin{center}
    \epsfig{scale=.35,file=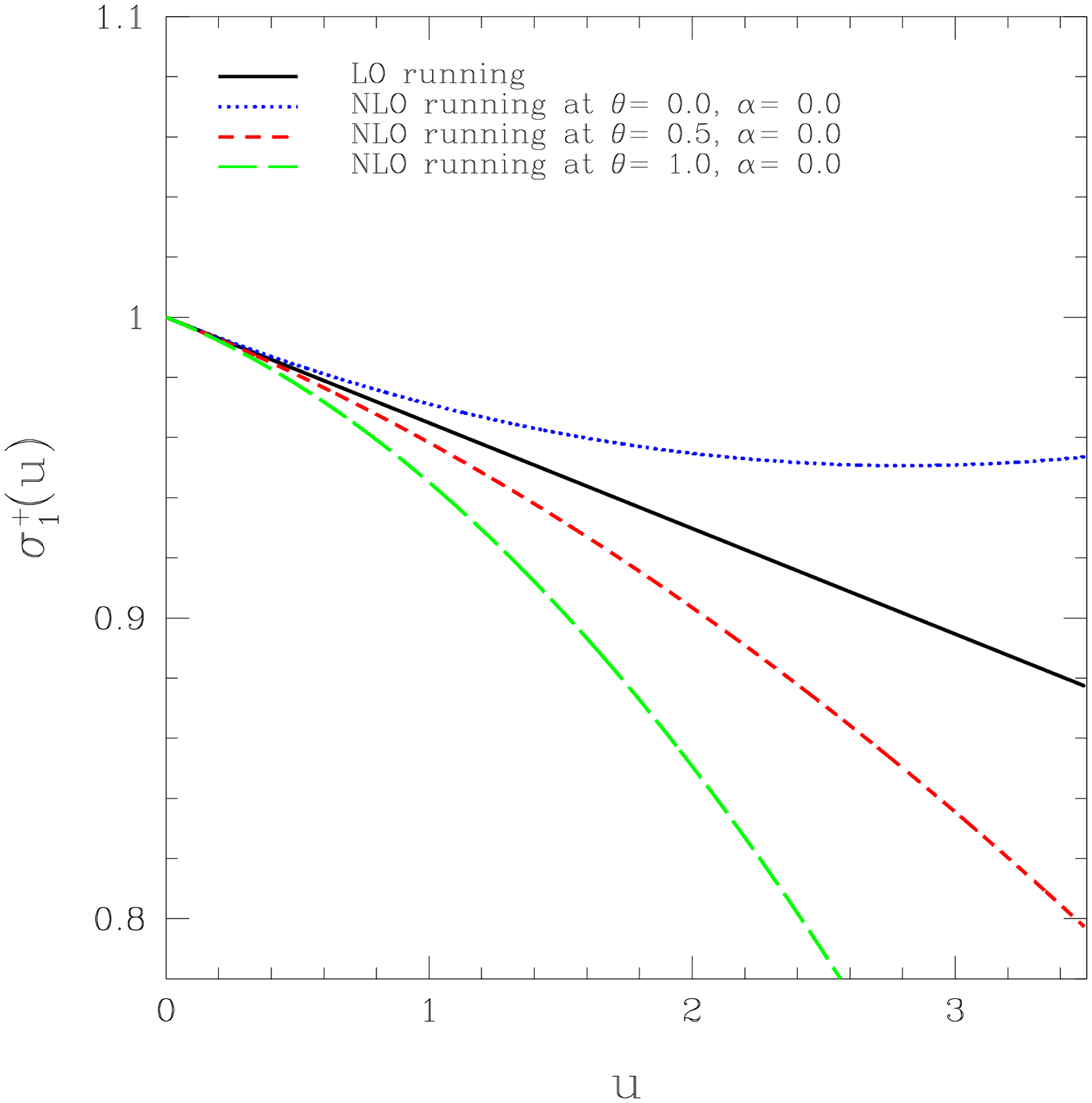}
      \end{center}
    \end{minipage}
    \begin{minipage}{.45\linewidth}
      \psfrag{xlabel}[c][bc][1][0]{$T/a$}
      \psfrag{ylabel}[c][c][1][0]{$\log_{10}r$}
      \begin{center}
    \epsfig{scale=.35,file=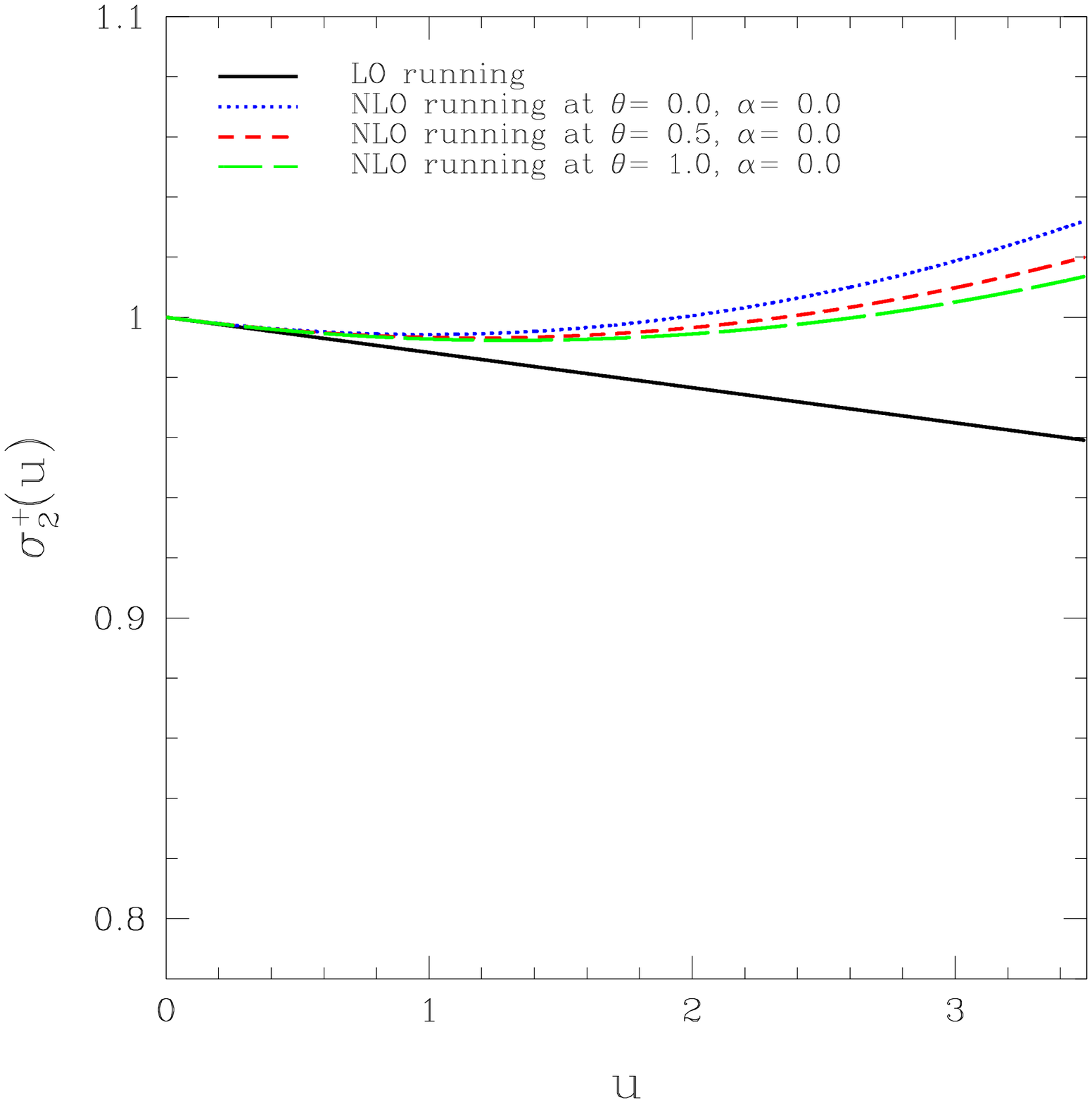}
      \end{center}
    \end{minipage}
    \caption{
      On the left(right) side the step
      scaling function of $\cQ_1'^+$($\cQ_2'^+$) at NLO 
      and $N_{\rm f}=0$ is plotted vs. the squared renormalised coupling in the
      SF scheme. The boundary sources choice is $s=1$,
      and the $\alpha$-parameter is set to zero.
      }
\label{fig_cutoff1}
  \end{center}
  \vspace{-.4cm}
  \label{figevol1}
\end{figure}

\begin{figure}[!ht]
  \begin{center}
    \begin{minipage}{.45\linewidth}
      \psfrag{xlabel}[c][bc][1][0]{$T/a$}
      \psfrag{ylabel}[c][c][1][0]{$\log_{10}r$}
      \begin{center}
    \epsfig{scale=.35,file=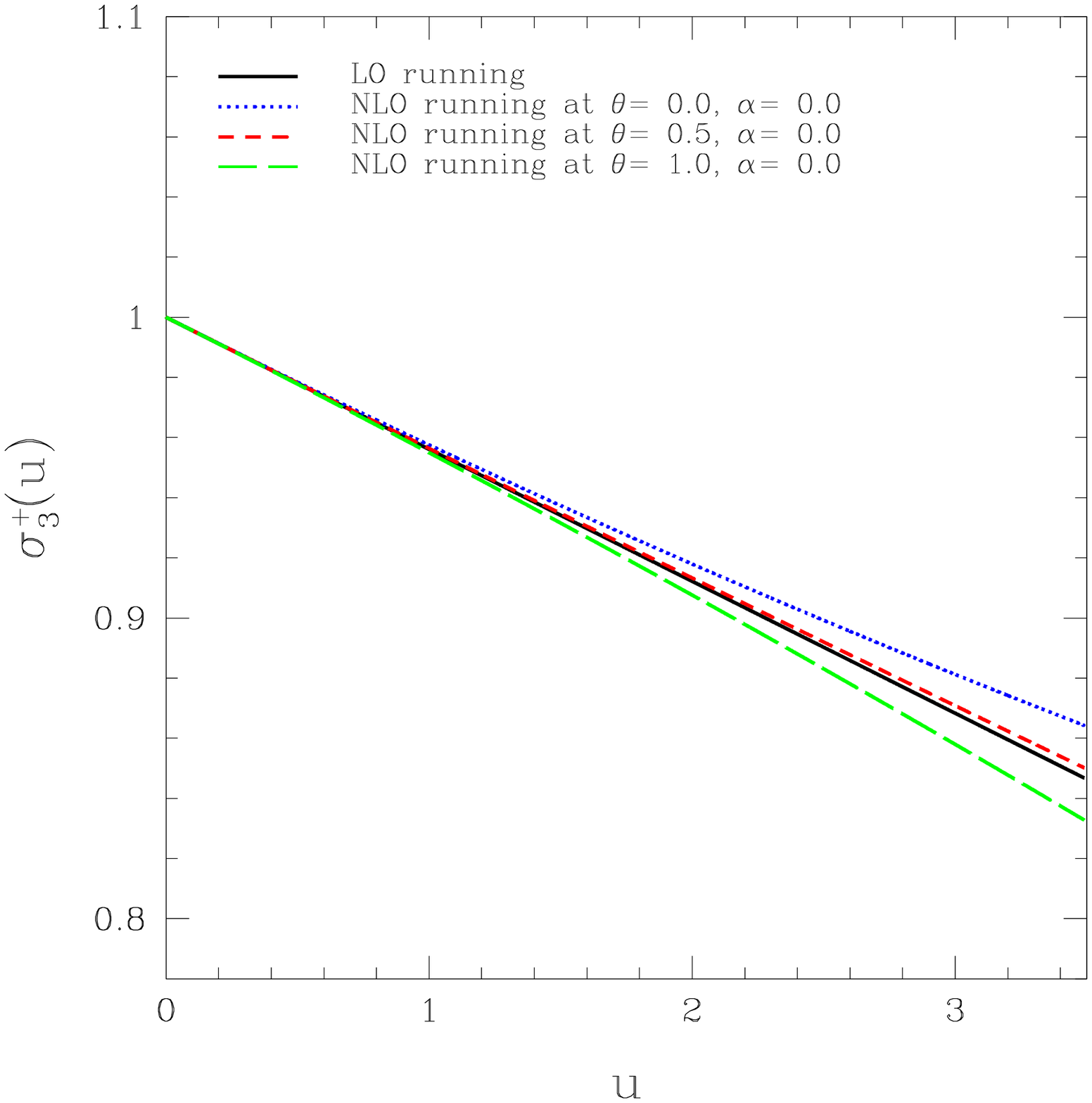}
      \end{center}
    \end{minipage}
    \begin{minipage}{.45\linewidth}
      \psfrag{xlabel}[c][bc][1][0]{$T/a$}
      \psfrag{ylabel}[c][c][1][0]{$\log_{10}r$}
      \begin{center}
    \epsfig{scale=.35,file=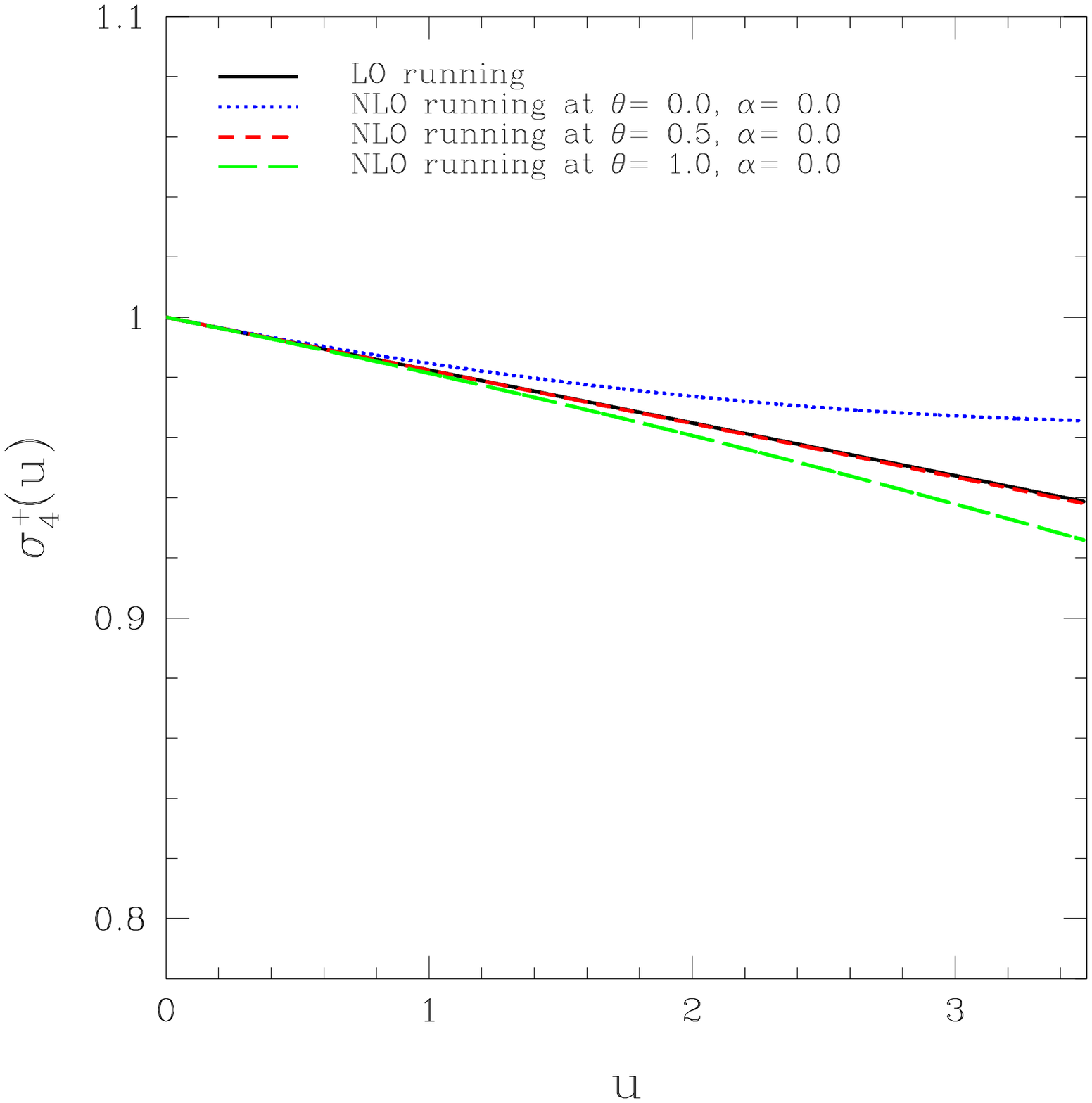}
      \end{center}
    \end{minipage}
    \caption{
      On the left(right) side the step
      scaling function of $\cQ_3'^+$($\cQ_4'^+$) at NLO 
      and $N_{\rm f}=0$ is plotted vs. the renormalised coupling in the
      SF scheme. The boundary sources choice is $s=1$,
      and the $\alpha$-parameter is set to zero.
	  }
\label{fig_cutoff2}
  \end{center}
  \vspace{-.4cm}
  \label{figevol2}
\end{figure}

The non-perturbative RG-evolution of the four-quark operators in the
diagonal basis (\ref{diagbasis}) is obtained through the computation
of the step-scaling functions
\begin{equation}
\label{stepscaling} \sigma_{k;\alpha}^{\lp;(s)}(u) = \lim_{a\to
0}\Sigma_{k;\alpha}^{\lp;(s)}(u,a/L)\ ,\qquad
\Sigma_{k;\alpha}^{\lp;(s)}(u,a/L)) =
\frac{\cZ_{k;\alpha}'^{+;(s)}(g_0,a/2L)}{\cZ_{k;\alpha}'^{+;(s)}(g_0,a/L)}\biggr|_{\bar
g^2(L)=u}\ .
\end{equation}
These ratios of renormalisation constants provide the operator running
between the scales $\mu=1/L$ and $\mu=1/2L$. The advantage of
introducing such ratios is related to the compensation of logarithmic
divergences between numerator and denominator, thus resulting in a
finite continuum limit. Cutoff effects can be therefore completely
decoupled from the continuum RG-evolution. We are concerned here with
the perturbative expansion of (\ref{stepscaling}) in the renormalised
coupling, that is
\begin{equation}
\label{pertevol} \sigma_{k;\alpha}^{\lp;(s)}(u) = 1 +
\sigma_{k;\alpha}^{\lp;(s,1)}u + \sigma_{k;\alpha}^{\lp;(s,2)}u^2 +
{\rm O}(u^3)\, .
\end{equation}
The first two terms of this expansion depend upon the LO and NLO 
anomalous dimension coefficients. They read explicitely
\begin{align}
\sigma_{k;\alpha}^{\lp;(s,1)} & = \gamma_{k}'^{\lp;(0)}\ln 2\ , \nonumber \\
\sigma_{k;\alpha}^{\lp;(s,2)} & = \gamma_{k;\rm SF}'^{\lp;(1)}\ln 2 +
\left[\frac{1}{2}(\gamma_{k}'^{\lp;(0)})^2 +
b_0\gamma_{k}'^{\lp;(0)}\right](\ln 2)^2\, .
\end{align}
A graphical representation of (\ref{pertevol}) for the whole operator
basis in some particular cases is provided by the four plots of
Figures~\ref{fig_cutoff1} and~\ref{fig_cutoff2}, in the range of
values of $g^2_{\rm SF}$ used in previous non-perturbative studies by the
ALPHA Collaboration.


The rate of convergence of the step-scaling functions
toward the continuum limit at LO can be expressed in terms of the first
non-trivial coefficient of the perturbative expansion (analogous to
(\ref{pertevol})) of $\Sigma_{k;\alpha}^{\lp;(s)}(u,a/L)$ via the ratio
\begin{equation}
\label{cutoffsigma}
\delta_{k;\alpha}^{\lp;(s)}(a/L) = \frac{\Sigma_{k;\alpha}^{\lp;(s,1)}(a/L) - \sigma_{k;\alpha}^{\lp;(s,1)}}{
\sigma_{k;\alpha}^{\lp;(s,1)}}\ ,
\end{equation}
where
\begin{equation}
\Sigma_{k;\alpha}^{\lp;(s,1)}(a/L) = \cZ_{k;\alpha}'^{+;(s,1)}(2L/a) - 
\cZ_{k;\alpha}'^{+;(s,1)}(L/a)\ .
\end{equation}
In order to compare the perturbative lattice artefacts
(\ref{cutoffsigma}) with the ones obtained from the corresponding
non-perturbative Monte Carlo simulations, the same definition of the
critical mass, based on the PCAC Ward identity, should be
adopted. This point has been extensively explained in
\cite{Palombi:2005zd}, where the numerical values of $am_c^{(1)}(L/a)$
from $L/a=6$ to $L/a=32$ have been provided ({\it cf.} Table~3 in that
work). That discussion will not be repeated here. Since our codes ran
up to $L/a=48$, we are in the position to extend the aforementioned
table to include the additional points. The new numbers are reported
in Table~\ref{tab:mcrit}.
\TABLE[!h]{ \centering
  \begin{tabular}{ccc}
    \\\hline\hline\\[-2.0ex]
    $L/a$ & $am_c^{(1)}(L/a)|_{c_{\rm sw}=1}/C_{\rm F}$ & $am_c^{(1)}(L/a)|_{c_{\rm sw}=0}/C_{\rm F}$ \\[0.5ex]
    \hline\\[-1.0ex]
    34 & -0.20255637783  & -0.32544080501  \\
    36 & -0.20255639414  & -0.32547023220  \\
    38 & -0.20255640819  & -0.32549515390  \\
    40 & -0.20255642028  & -0.32551644442  \\
    42 & -0.20255643068  & -0.32553477599  \\
    44 & -0.20255643965  & -0.32555067224  \\
    46 & -0.20255644740  & -0.32556454600  \\
    48 & -0.20255645412  & -0.32557672619  \\[1.0ex]
    \hline\hline
  \end{tabular}
  \caption{The one-loop coefficients of the critical mass as obtained from the PCAC
    Ward Identity at finite lattice size. For the parameter choices made here, the convergence to the values
  at infinite lattice size is quadratic/cubic in $(a/L)$, for standard/${\rm O}(a)$ improved Wilson quarks.}
  \label{tab:mcrit}
}

\begin{figure}[!ht]
  \begin{center}
    \epsfig{scale=.55,file=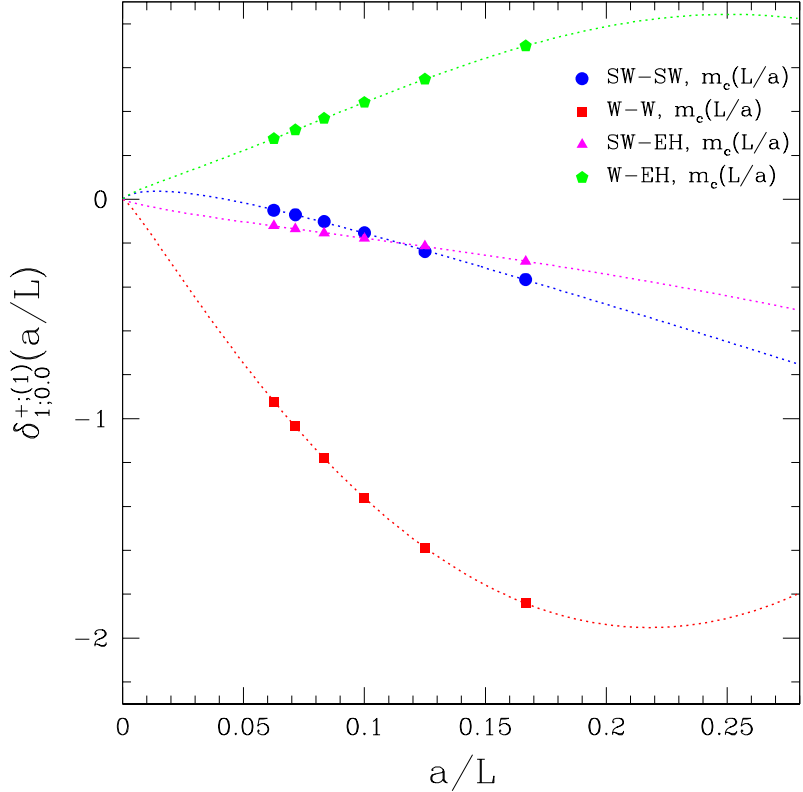}
  \end{center}
  \caption{\label{fig_artefacts} Comparison between cutoff effects of
  the step scaling function of the static-light and light-light versions
  of the four-quark
  operator $\cQ_1'^\lp$. Light quarks are described in terms of the
  unimproved (W) or improved (SW) Wilson action. Static quarks are
  always described by the Eichten-Hill (EH) action. The critical 
  mass has been obtained from the PCAC Ward Identity.}
  \vspace{-.4cm}
  \label{figcutoff}
\end{figure}

In practice, non-perturbative simulations based on the Eichten-Hill
discretisation of the heavy quark fields should better be avoided,
given the bad intrinsic signal-to-noise ratio~(\ref{EHaction},
\ref{eq_latt_der})~\cite{DellaMorte:2005yc}. Nevertheless, it is
instructive to compute lattice artefacts in perturbation theory for
the Eichten-Hill action, if only to check whether the use of static
fields enhances lattice artefacts with respect to the purely
relativistic case. A comparison between static-light and light-light
four-quark operators in a typical situation is shown in
Figure~\ref{fig_artefacts}. All data refer to $\cQ_1^\lp$, 
employing a renormalisation scheme in which the boundary sources have
a Dirac structure $[\gamma_5,\gamma_5,\gamma_5]$ and where the
normalisation of the four-quark correlator involves only $f_1^{ll}$. 
Relativistic data are taken from \cite{Palombi:2005zd}. Taken at face 
value, the plot leads to the conclusion that the light quark action 
is the main responsible for the presence of relatively large lattice 
artefacts: once this has been chosen, static-light and light-light 
four-quark operators come to be affected by cutoff effects of a 
similar size.

%% file: mapping.tex
\section{tmQCD for $B^0$--$\bar B^0$ mixing}
\label{sec:mapping}

As stated above, our interest in the renormalisation of $\cQ^+_k$
stems from the fact that the physical $\Delta B=2$ matrix elements
involving the operators $Q^+_1,Q^+_2$, with
$\psi_1=\psi_2\equiv\psi_\ell$ (where in practice the flavour label
$\ell$ denotes either a $d$ or an $s$ quark), can be mapped onto
matrix elements of $\cQ^+_1,\cQ^+_2$ computed in some suitable tmQCD
regularisation. This reduces to a minimum the uncertainties related to
operator renormalisation, which takes place essentially in the same
way as if exact chiral symmetry were present. Now we will construct
specific tmQCD regularisations which realize this mapping. This
technique is a generalisation of the ones already developed for
$B_K$~\cite{Frezzotti:2000nk,Guagnelli:2005zc}.

We will work in the so-called ``twisted basis'', and concentrate
first on the case of the $B_B$ parameter where $\psi_\ell=d$. Let us
consider a quark doublet $\psi^T = (u,d)$, for which we specify the
tmQCD action
\begin{gather}
\label{tmQCDaction}
S^{\rm tmQCD}[\psi,\bar\psi] = a^4\sum_{x}\
      \left\{\bar\psi(x)\left[D_{\rm w} + m +
	i\mu\tau^3\gamma_5\right]\psi(x)\right\} \, ,
\end{gather}
where $D_{\rm w}$ is the usual Wilson-Dirac operator (with or without
a SW term). The choice of action for the other relativistic quark
flavours is immaterial to the argument. The equivalence of tmQCD to
standard QCD has been first established in
\cite{Frezzotti:2000nk}. Given a multi-local gauge-invariant operator
$\cO(x_1,\dots,x_n)$, the equivalence amounts to the identity between
renormalised correlation functions
\begin{equation}
\label{tmQCDinv}
\langle \tilde \cO_{\rm R}(x_1,\dots,x_n)\rangle_{(M_{\rm R},0)} = \langle
\cO_{\rm R}(x_1,\dots,x_n)\rangle_{(m_{\rm R},\mu_{\rm R})}\ ,
\end{equation}
which holds in the regularised theory up to cutoff effects, and is
exact in the continuum limit. In the above expression, the relation
between the operators $\tilde\cO$ and $\cO$ is provided by the axial
rotation of the quark fields which relates QCD to tmQCD, viz.
\begin{gather}
\label{tmQCDrot}
\begin{split}
  \tilde\psi &= \exp\left(i\alpha\gamma_5\tau^3/2\right)\psi
     \\ 
     \tilde{\bar\psi} &=
    \bar\psi\exp\left(i\alpha\gamma_5\tau^3/2\right)\ ,
  \end{split}
\end{gather}
where the twist angle $\alpha$ is defined in terms of the renormalised
mass parameters of the tmQCD action as
\begin{gather}
\tan(\alpha) = \frac{\mu_{\rm R}}{m_{\rm R}} \, ,
\end{gather}
and the physical renormalised quark mass $M_{\rm R}$ is given by
\begin{gather}
M_{\rm R} = \sqrt{m_{\rm R}^2+\mu_{\rm R}^2} \, .
\end{gather}
Let us now consider the static-light four-quark operators $\cO^\lp_{\rm VV+AA}$ and
$\cO^\lp_{\rm VA+AV}$ with light flavours $\psi_1=\psi_2=d$.
We observe that the rotation (\ref{tmQCDrot}) implies
\begin{equation}
  \label{VVAArotVAAV}
  \tilde\cO^\lp_{\rm VV+AA} =
  \cos(\alpha)\cO^\lp_{\rm VV+AA} -
  i\sin(\alpha)\cO^\lp_{\rm VA+AV}\ .
\end{equation}
In particular, at $\alpha=\pi/2$, which is known as the {\it maximally
  twisted} case, (\ref{VVAArotVAAV}) simplifies to 
\begin{equation}
  \label{fulltwistrot}
  \tilde\cO^\lp_{\rm VV+AA} =
  -i\cO^\lp_{\rm VA+AV}\ .
\end{equation}
Exactly the same property holds for the operator ${\rm SS+PP}$, for which
one finds again
\begin{equation}
  \tilde\cO^\lp_{\rm SS+PP} =
  -i\cO^\lp_{\rm SP+PS}\ .
\end{equation}
This demonstrates explicitly that the matrix elements of $\cO^\lp_{\rm
VV+AA}$ and $\cO^\lp_{\rm SS+PP}$, responsible for the particle mixing in
the SM and within the static approximation of QCD, can be obtained
from a computation of the matrix element of $\cO^\lp_{\rm VA+AV}$ and
$\cO^\lp_{\rm SP+PS}$ in tmQCD at $\alpha=\pi/2$. Since in mass independent 
schemes, such as the SF, all dependence of renormalisation factors on 
the mass parameters drops out, it is clear that tmQCD does not 
spoil the renormalisation pattern of the operator basis (\ref{diagbasis}). 
In particular, the combinations 
$\cQ_1'\equiv \cO^\lp_{\rm VA+AV}$  and 
$\cQ_2'\equiv \cO^\lp_{\rm VA+AV} + 4\cO^\lp_{\rm SP+PS}$ 
renormalise purely multiplicatively.

In case one is interested in the $B_s^0$--$\bar B_s^0$ mixing
amplitude, it is enough to maximally twist a quark doublet which
contains the $s$ quark, e.g. $(c,s)$, the action for which would read
exactly as eq.~(\ref{tmQCDaction}), save for the eventual introduction
of non-degenerate masses for the two quarks of the doublet, along the
lines of~\cite{Pena:2004gb}. The action for a twisted $(c,s)$ doublet
would then read
\begin{gather}
\label{tmQCDaction_2}
S^{\rm tmQCD}[\psi,\bar\psi] = a^4\sum_{x}\
      \left\{\bar\psi(x)\left[D_{\rm w} + \mathbf{m} +
	i\boldsymbol{\mu}\gamma_5\right]\psi(x)\right\}
\end{gather}
with $\mathbf{m}={\rm diag}(m_c,m_s)$ and $\boldsymbol{\mu}={\rm diag}(\mu_c,\mu_s)$,
and the constraint
\begin{gather}
\tan(\alpha) = -\frac{\mu_{s,{\rm R}}}{m_{s,{\rm R}}}= \frac{\mu_{c,{\rm R}}}{m_{c,{\rm R}}} \, .
\end{gather}
A potential shortcoming of eq.~(\ref{tmQCDaction_2}) comes about in
case it is taken as the action for dynamical $c,s$ quarks, since
$M_{s,{\rm R}} \ne M_{c,{\rm R}}$ would then induce a phase in the
fermion determinant. One may then consider more sophisticated chiral
rotations that keep the determinant real, as
in~\cite{Frezzotti:2003xj}. If $c,s$ are kept quenched, or interpreted
as valence quarks, no such subtlety arises.

We conclude that the use of suitable tmQCD regularisations avoids the
need of determining mixing coefficients for the renormalisation of the
matrix elements entering the $B^0$--$\bar B^0$ amplitude in the static
approximation. For an alternative analysis of operator mixing using a
different tmQCD regularisation, we refer the reader to
ref.~\cite{DellaMorte:2004wn}.

An additional advantage brought in by the use of maximally tmQCD
is the automatic $\Oa$ improvement of bare matrix elements of the
above four-fermion operators.
This property does not hold, on the other hand, for the renormalisation constants computed
within the SF schemes discussed in previous sections. In order to obtain O($a$) improved
renormalization constants one should use modified SF schemes, as proposed
in refs.~\cite{Sint:2005qz,Frezzotti:2005zm}.  

In order to show that the bare matrix elements are automatically O($a$)
improved we extend the argument in Appendix A of
ref.~\cite{Frezzotti:2005gi}. The first observation is that, at maximal 
twist, the only O($a$) counterterms to the static action are 
proportional to the dimension five operators
\begin{gather}
\tr(\boldsymbol{\mu}^2)\,
(\bar{\psi}_h\psi_h+\bar{\psi}_{\bar h}\psi_{\bar h}) \, ,~~~~~~~~~~
(\tr(\boldsymbol{\mu}))^2\,
(\bar{\psi}_h\psi_h+\bar{\psi}_{\bar h}\psi_{\bar h}) \, .
\end{gather}
(In the simple case $\boldsymbol{\mu}=\mu\tau^3$ 
there is one single counterterm proportional to 
$\mu^2(\bar{\psi}_h\psi_h+\bar{\psi}_{\bar h}\psi_{\bar h})$.)
These counterterms merely generate a shift of the static quark
self-energy~\footnote{Moreover, they are obviously absent in the
quenched approximation.}. The second observation is that all the 
possible O($a$) (dimension seven) counterterms will have the same static field
content of the original (dimension six) four-fermion operator, and
will differ from it only by 
the addition of mass factors or derivatives. It is then possible to extend
the symmetry ${\cal P}\times{\cal D}_d\times(\mu\to-\mu)$ of the relativistic
action~\cite{Frezzotti:2005gi} (where $\cal P$ is the physical parity and
${\cal D}_d$ is defined in ref.~\cite{Frezzotti:2003ni}) to
include static quarks. ${\cal P}$ and ${\cal D}_d$
will now be defined to be 
\begin{eqnarray}  
{\cal{P}}:\left \{\begin{array}{lll} 
U_0(x)&\rightarrow &U_0(x^\pi)\\
U_k(x)&\rightarrow &U_k^{\dagger}(x^\pi-a\hat{k})\\
\psi(x)&\rightarrow & i\gamma_0\gamma_5\tau_3\psi(x^\pi)\\
\bar{\psi}(x)& \rightarrow &\bar{\psi}(x^\pi)\tau_3\gamma_5\gamma_0i\\ 
\psi_h(x)&\rightarrow & \psi_h(x^\pi)\\
\bar{\psi}_h(x)& \rightarrow &\bar{\psi}_h(x^\pi)\\ 
\psi_{\bar h}(x)&\rightarrow & -\psi_{\bar h}(x^\pi)\\
\bar{\psi}_{\bar h}(x)& \rightarrow &-\bar{\psi}_{\bar h}(x^\pi) 
\end{array}\right . & \qquad \qquad & 
{\cal{D}}_d : \left \{\begin{array}{lll}     
U_\mu(x)&\rightarrow & U_\mu^\dagger(-x-a\hat\mu) \\
\psi(x)&\rightarrow & -i \psi(-x)  \\
\bar{\psi}(x)&\rightarrow & -i \bar{\psi}(-x)\\  
\psi_h(x)&\rightarrow & \psi_{\bar h}(-x)  \\
\bar{\psi}_h(x)&\rightarrow & \bar{\psi}_{\bar h}(-x)\\  
\psi_{\bar h}(x)&\rightarrow & \psi_{h}(-x)  \\
\bar{\psi}_{\bar h}(x)&\rightarrow & \bar{\psi}_{h}(-x)  
\end{array}\right . \label{FIELDT} 
\end{eqnarray} 
where $(x_0,\bx)^\pi = (x_0,-\bx)$. Using ${\cal P}\times{\cal D}_d\times(\mu\to-\mu)$ 
one immediately concludes that all the relevant dimension seven operators 
have opposite parity with respect to the dimension six ones. Using
the same arguments of ref.~\cite{Frezzotti:2005gi}, one then 
concludes that no O($a$) appears in
the Symanzik expansion of the relevant correlation functions.

%% file: conclusions.tex
\section{Conclusions \label{sec:concl}}

In this paper we have shown that the renormalisation problem of
heavy-light four-quark operators in the static approximation can be
tackled for Wilson-like fermions without the need to perform finite
subtractions.

Owing to the presence of static quark fields, the flavour switching
symmetries, which in the relativistic case have proved so useful
\cite{Bernard:1987pr,Donini:1999sf} for imposing constraints on the
mixing, are very much reduced. However, this lack is compensated by
the heavy quark symmetry, spatial rotations and a set of discrete
symmetries, such as time reversal. The emerging renormalisation and
mixing pattern is then quite similar to the relativistic theory: while
chiral symmetry breaking generated by the Wilson term induces mixing
among different chiralities of parity-even operators in the lattice
regularised theory, such mixings are completely absent in the
parity-odd sector.

Twisted-mass QCD can be used to relate the operator bases in the
parity-even and parity-odd sectors also in the static
approximation. In particular, we have shown how to do this for the 
operators that contribute to $B^0$--$\bar B^0$ mixing in the Standard 
Model, using a maximal twist setup that brings in, as a bonus, the 
potential for automatic $\Oa$ improvement.

A fully non-perturbative determination of the renormalisation factors
of four-quark operators in the framework of the Schr\"odinger
functional appears entirely feasible at this point, provided that one
can overcome the well-known problem of the Eichten-Hill action, namely
the exponential growth of statistical fluctuations at large Euclidean
times \cite{Palombi:npren}. Here the hope is that the methods described in
refs. \cite{DellaMorte:2003mn,DellaMorte:2005yc} turn out to be as
useful as in the simpler case of heavy-light bilinears.

We have verified explicitly the expected mixing pattern in an
extensive perturbative calculation at one loop. Thereby we have also
obtained the NLO anomalous dimensions, which will be an important
ingredient in future non-perturbative determinations of the
renormalisation factors. Furthermore, our perturbative calculation
can be used to optimise the choice of renormalisation prescription in the
forthcoming numerical simulations.

As we have mentioned above, at the level of $B^0$--$\bar B^0$ amplitudes the
matching between HQET and QCD requires to compute matrix elements of the
operators $\cO^+_{\rm VV+AA}$ and $\cO^+_{\rm SS+PP}$, which are mapped
via tmQCD onto the operators $\cO^+_{\rm VA+AV}$ and $\cO^+_{\rm SP+PS}$. 
Therefore, as far as renormalisation is concerned, one is then
faced with the task of computing the step-scaling functions for
the relevant pair of operators, which renormalise
multiplicatively, i.e. $\cO^+_{\rm VA+AV}$ and $\cO^+_{\rm VA+AV} 
+4\cO^+_{\rm SP+PS}$.

The static approximation considered in this work only represents the
lowest order of HQET, and hence all results for phenomenologically
relevant quantities are subject to corrections in powers of the
inverse heavy quark mass. While there are strategies in place which
are designed for determining the leading $1/M$ corrections
non-perturbatively \cite{Heitger:2003nj}, it is also possible to
interpolate lattice results between the static approximation and the
regime of relativistic quarks with masses around that of the charm
quark. Our findings may serve to obtain high-precision results for
$B^0- \bar B^0$ mixing amplitudes in the static approximation, which
in turn are required to perform reliable interpolations to the
physical $b$-quark mass.

%% file: acknow.tex
\subsection*{Acknowledgements}

We are grateful to J.~Heitger and R.~Sommer for their participation in
the early stages of this work. We thank D.~Be\'cirevi\'c, M.~Della
Morte, F.~Mescia and especially J.~Reyes for useful
discussions. F.P.~acknowledges the Alexander-von-Humboldt Stiftung for
financial support. Hospitality offered by CERN (F.P., M.P.) and DESY
(C.P.) during the preparation of this work is thankfully acknowledged.


%% file: appendA.tex
\appendix

\section[Appendix A]{Constraints from heavy quark spin symmetry and
$H(3)$ spatial rotations on the mixing pattern \label{app:A}}

We now describe the procedure followed to impose the constraints from
heavy quark spin symmetry and cubic rotations. It applies
identically to both the parity-even and the parity-odd sectors and we
choose to describe it for the latter. It turns out that, in this
particular case, there is no need for considering a maximal set of
independent symmetry transformations, because the final constraints
are already obtained by considering a finite spin transformation of
the heavy fields, e.g.
\begin{align}
\label{hqss}
\Phi_1:\qquad& \bar\psi_{h} \to
\bar\psi_{h}\gamma_2\gamma_3\ , \quad  \bar\psi_{\bar h} \to
\bar\psi_{\bar h}\gamma_2\gamma_3\ , 
\end{align}
and two lattice spatial rotations, e.g. 
\begin{align}
\label{H3}
\Phi_2:\qquad\cR(\ \hat 1 \to \hat 2\ ) \ \mbox{rotates the $\hat 1$ axis
      onto the $\hat 2$ axis}, \nonumber \\[1.5ex]
\Phi_3:\qquad\cR(\ \hat 2 \to \hat 3\ ) \ \mbox{rotates the $\hat 2$ axis
      onto the $\hat 3$ axis,} 
\end{align}
alone. The subspace spanned by~(\ref{basis}) is not invariant under
the set of transformations (\ref{hqss}). We hence give up temporarily
Lorentz invariance (which, as we will see, will be recovered
naturally) and consider an enlarged basis containing eight operators,
\begin{equation}
\label{newbasis}
\cO^\pm =
(\cO^\pm_{V_0A_0+A_0V_0},\dots,\cO^\pm_{V_3A_3+A_3V_3},\cO^\pm_{V_0A_0-A_0V_0},\dots,
\cO^\pm_{V_3A_3-A_3V_3})^T\ ,
\end{equation}
which can generate, when properly combined, the original parity-odd
basis (\ref{basis})~\footnote{Notice that, due the constraints
(\ref{constraints}), the operators $\cO^\pm_{\rm SP+PS}$ and
$\cO^\pm_{\rm SP-PS}$ are contained in the basis~(\ref{newbasis})}. The
analysis performed in section~\ref{sec:mixing} by using chiral
symmetry can be carried over to (\ref{newbasis}), which can be
accordingly shown to renormalise as \begin{equation}
\label{rennewbasis}
\cO^\pm_R = z^{\pm}(\mathds{1} + \delta^{\pm})\cO^\pm\ .
\end{equation}
Here $z^\pm$ are block diagonal matrices containing two $(4\times 4)$
scale-dependent blocks while $\delta^\pm$ are block off-diagonal
matrices, containing two $(4\times 4)$ scale-independent blocks. The
advantage of using (\ref{newbasis}) is that the new basis is closed
under (\ref{hqss}) and (\ref{H3}), and the matrices $\Phi_k$ that
implement the symmetry rotations can be constructed
explicitly. Moreover, it should be observed that in order to preserve
the renormalisation structure determined by chiral symmetry, the two
matrices $z^\pm$ and $\delta^\pm$ have to satisfy the symmetry
constraints independently, namely
\begin{align}
z^\pm & = \Phi_k z^\pm\Phi_k^{-1}\ ,\nonumber \\[1.5ex]
\delta^\pm & = \Phi_k\delta^\pm\Phi_k^{-1}\ .
\end{align}
The explicit form of the matrices $\Phi_k$ is easily found out to be:
\begin{gather}
\Phi_1 =
\left(\begin{array}{rrrrrrrr}
  0 &  1 &  0 &  0 &  0 &  0 &  0 &  0 \\
  1 &  0 &  0 &  0 &  0 &  0 &  0 &  0 \\
  0 &  0 &  0 &  1 &  0 &  0 &  0 &  0 \\
  0 &  0 &  1 &  0 &  0 &  0 &  0 &  0 \\
  0 &  0 &  0 &  0 &  0 & -1 &  0 &  0 \\
  0 &  0 &  0 &  0 & -1 &  0 &  0 &  0 \\
  0 &  0 &  0 &  0 &  0 &  0 &  0 &  1 \\
  0 &  0 &  0 &  0 &  0 &  0 &  1 &  0 \\
\end{array}\right) \ ,\quad
\Phi_2 =
\left(\begin{array}{rrrrrrrr}
  1 &  0 &  0 &  0 &  0 &  0 &  0 &  0 \\
  0 &  0 &  1 &  0 &  0 &  0 &  0 &  0 \\
  0 &  1 &  0 &  0 &  0 &  0 &  0 &  0 \\
  0 &  0 &  0 &  1 &  0 &  0 &  0 &  0 \\
  0 &  0 &  0 &  0 &  1 &  0 &  0 &  0 \\
  0 &  0 &  0 &  0 &  0 &  0 &  1 &  0 \\
  0 &  0 &  0 &  0 &  0 &  1 &  0 &  0 \\
  0 &  0 &  0 &  0 &  0 &  0 &  0 &  1 \\
\end{array}\right) \ ,\quad
\Phi_3 =
\left(\begin{array}{rrrrrrrr}
  1 &  0 &  0 &  0 &  0 &  0 &  0 &  0 \\
  0 &  1 &  0 &  0 &  0 &  0 &  0 &  0 \\
  0 &  0 &  0 &  1 &  0 &  0 &  0 &  0 \\
  0 &  0 &  1 &  0 &  0 &  0 &  0 &  0 \\
  0 &  0 &  0 &  0 &  1 &  0 &  0 &  0 \\
  0 &  0 &  0 &  0 &  0 &  1 &  0 &  0 \\
  0 &  0 &  0 &  0 &  0 &  0 &  0 &  1 \\
  0 &  0 &  0 &  0 &  0 &  0 &  1 &  0 \\
\end{array}\right) \ .\nonumber
\end{gather}
Once all the constraints are imposed one gets~\footnote{To simplify the
notation we neglect the superscript $^\pm$ on $z_1$, $z_2$, $z'_1$, $z'_2$,
$\delta_1$, $\delta_2$, $\delta'_1$, $\delta'_2$ despite the fact that
these matrix elements are in general different in the ${\cal S} =
\pm1$ sectors.}:
\begin{gather}
\nonumber
z^\pm =
\left(\begin{array}{rrrrrrrr}
  z_1 &  z_2 &  z_2 &  z_2 &  0 &  0 &  0 &  0 \\
  z_2 &  z_1 &  z_2 &  z_2 &  0 &  0 &  0 &  0 \\
  z_2 &  z_2 &  z_1 &  z_2 &  0 &  0 &  0 &  0 \\
  z_2 &  z_2 &  z_2 &  z_1 &  0 &  0 &  0 &  0 \\
  0 &  0 &  0 &  0 &  z'_1 & -z'_2 & -z'_2 & -z'_2 \\
  0 &  0 &  0 &  0 & -z'_2 &  z'_1 &  z'_2 &  z'_2 \\
  0 &  0 &  0 &  0 & -z'_2 &  z'_2 &  z'_1 &  z'_2 \\
  0 &  0 &  0 &  0 & -z'_2 &  z'_2 &  z'_2 &  z'_1 \\
\end{array}\right) \ ,\quad
\delta^\pm =
\left(\begin{array}{rrrrrrrr}
  0 &  0 &  0 &  0 & -\delta_1 &  \delta_2 &  \delta_2 &  \delta_2 \\
  0 &  0 &  0 &  0 & -\delta_2 &  \delta_1 &  \delta_2 &  \delta_2 \\
  0 &  0 &  0 &  0 & -\delta_2 &  \delta_2 &  \delta_1 &  \delta_2 \\
  0 &  0 &  0 &  0 & -\delta_2 &  \delta_2 &  \delta_2 &  \delta_1 \\
 -\delta'_1 & -\delta'_2 & -\delta'_2 & -\delta'_2 &  0 &  0 &  0 &  0 \\
  \delta'_2 &  \delta'_1 &  \delta'_2 &  \delta'_2 &  0 &  0 &  0 &  0 \\
  \delta'_2 &  \delta'_2 &  \delta'_1 &  \delta'_2 &  0 &  0 &  0 &  0 \\
  \delta'_2 &  \delta'_2 &  \delta'_2 &  \delta'_1 &  0 &  0 &  0 &  0 \\
\end{array}\right) \ .
\end{gather}
Now we go to a basis such that $z^\pm$ is diagonal. A convenient
choice is:
\begin{gather}
\{\cQ^\pm_1,\cQ^\pm_1+4\cQ^\pm_2,R_1^\pm,R_2^\pm,\cQ^\pm_3+2\cQ^\pm_4,\cQ^\pm_3-2\cQ^\pm_4,R_3^\pm,R_4^\pm\} \ ,
\end{gather}
where $R_j^\pm$ are some Lorentz non-invariant operators, the precise
expression of which is irrelevant for the rest of the argument. By
transforming into this basis it turns out that also the nontrivial
blocks of $\delta^\pm$ are diagonal\footnote{The triplet
$\{\cQ^\pm_1+4\cQ^\pm_2,R_1^\pm,R_2^\pm\}$ corresponds to an
eigenvalue with a three-dimensional associated subspace in the space
of operators, and so does $\{\cQ^\pm_3-2\cQ^\pm_4,R_3^\pm,R_4^\pm\}$
-- both for $z^\pm$ and $\delta^\pm$.}. We have therefore arrived to
the final conclusion that the basis~(\ref{diagbasis}) is closed under
renormalisation, with the mixing pattern presented
in~(\ref{diagmixing}).

%% file: appendB.tex
\section[Appendix B]{Integrals in the infinite volume theory
  \label{app:B}} 

The lattice contributions to the matching coefficients
$\cX^{(1)}_{{\rm ref,lat}}$, commonly expressed in terms of a basic
set of Feynman integrals in the momentum-space representation, have
been calculated and cross-checked by independent authors
\cite{Borrelli:1992fy,Gimenez:1998mw,Eichten:1989kb,DiPierro:1998ty}.
In all cases, their evaluation has been pursued through Monte Carlo
simulations (VEGAS), resulting in an average numerical precision of
three digits. On the other hand, the matching constants $\cX_{\rm SF,
lat}^{(1)}$, which provide the connection between the SF and the
lattice scheme, have been calculated with a better accuracy, as
explained in section~\ref{sec:pert_exp}. As a consequence, the
uncertainty on the NLO coefficients of the SF anomalous dimensions is
dominated by the lack of precision in the infinite lattice
integrals. A possible way out would be running the Monte Carlo
algorithms on faster computers and wait long enough for a couple of
digits more. A more attractive alternative is to use an analytical
trick to improve the quality of the results, obtaining at the same
time some insight into the peculiar nature of the static lattice
integrals. As an example, we consider
\begin{align}
d_1 & =
\label{d1}
\frac{1}{\pi^2}\int_{-\pi}^{\pi}d^4k\biggl[-4\theta(1-k^2)\frac{1}{k^4}
  + \frac{1}{4\Delta_1\Delta_2}
  +\frac{3}{16}\frac{1}{\Delta_2}\biggr] , \nonumber \\[1.5ex]
\Delta_1 & = \sum_{\mu=1}^4\sin^2\frac{k_\mu}{2}\ , \qquad \Delta_2 =
\sum_{\mu=1}^4\sin^2k_\mu + 4\Delta_1^2\ .
\end{align}
The first term in the integral, which comes from the static
propagator, diverges logarithmically at $k=0$. This contribution is
compensated by an opposite divergence of the subsequent terms, which
brings the final result to a finite value, namely $d_1\simeq 5.46$. In
principle, $d_1$ could be regularised through the usual lattice
discretisation of the integration variables, 
\begin{equation}
k_\mu\ \rightarrow\ \frac{2\pi}{N}n_\mu\ ,\qquad -\frac{N}{2}<
n_\mu\le \frac{N}{2}\ .
\end{equation}
If it were not for the $\theta(1-k^2)$-function, the integral would be
expected to behave like an ordinary lattice integral, i.e. its
convergence to the continuum would be determined by the asymptotic
formula
\begin{equation}
\label{converg}
\tilde d_1(N) = d_1 + \frac{a_1}{N} + \frac{b_1}{N}\log(N) +
\frac{a_2}{N^2} + \frac{b_2}{N^2}\log(N) +
O\left(\frac{1}{N^3}\right)\ ,
\end{equation}
where $\tilde d_1$ represents the lattice version of $d_1$. The
$\theta$-function, which is non-zero inside a spherical domain, produces an
explicit breaking of the hyper-cubic $H(4)$ symmetry, thus perturbing
the convergence pattern. This effect can be better understood by
defining   
\begin{equation}
\label{Itilde}
I\equiv\int_{-\pi}^\pi d^4k\ \theta(1-k^2)\frac{1}{k^4} \quad
\longrightarrow \quad \tilde I(N) =
\frac{1}{\pi^2}\sum_{n_i:\ n^2\ne 0}\frac{\theta\left(\frac{N^2}{4\pi^2}-n^2\right)}{n^4}\
, \end{equation}
and observing that the number of the lattice points that lie inside a
3-sphere $\Sigma$ with radius $R_\Sigma=N/2\pi$ increases 
irregularly when $N\to\infty$. The alternating excess or deficit of
integration points gives rise to the oscillating behaviour
characterising the dashed curve of Figure~4. In order to smooth
$\tilde d_1$, we propose to regularise $\tilde I$ in a way that formally
restores the hyper-cubic symmetry. To this aim, we introduce an order
parameter $\Delta V_\Sigma$, which provides a measure of the
spherical symmetry breaking produced by the lattice discretisation,   
\begin{figure}[!t]
\begin{center}
\epsfig{figure=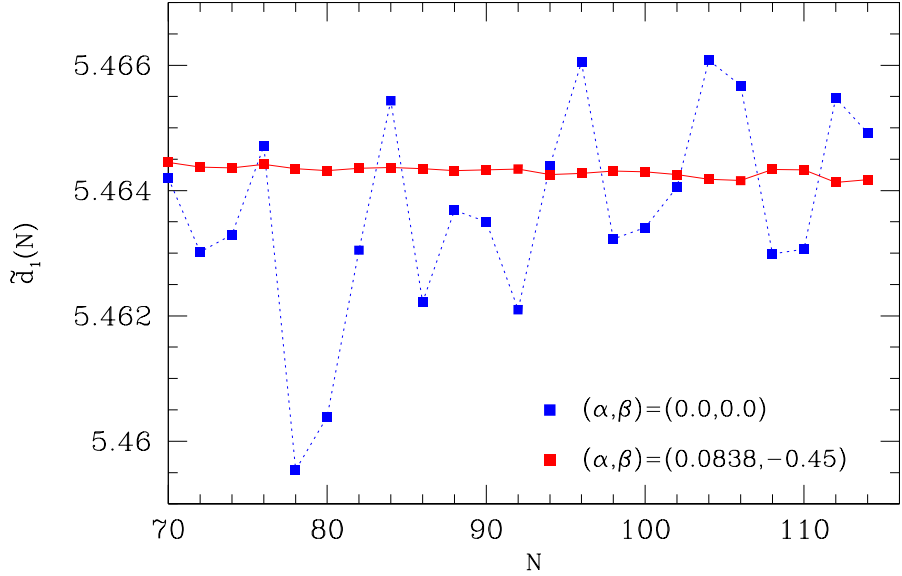, width=12.5 true cm}
\end{center}
\label{d1plot}
\vskip -5.0 true cm
\caption{Convergence pattern of $\tilde d_1(N)$.}
\end{figure}
\begin{equation}
\label{deltaS3}
\Delta V_\Sigma(N) = \frac{V_\Sigma - \tilde V_\Sigma(N)}{V_\Sigma}\ .
\end{equation}
Here $V_\Sigma = N^4/(32\pi^2)$ is the volume of the above-mentioned
3-sphere, and $\tilde V_\Sigma(N)$ represents the corresponding lattice
volume, obtained by just counting the number of the lattice points
that belong to the inner of $\Sigma$ at fixed $N$. The recovery of spherical
symmetry in the  continuum limit implies that $\Delta V_\Sigma$
vanishes when $N\to\infty$. In addition, $\Delta V_\Sigma$ describes
a surface effect at large values of $N$, and its rate of vanishing is 
therefore proportional to $1/N$, up to fluctuations. We now define  
\begin{equation}
\tilde I'(N) = \left\{ 1 + \alpha \Delta V_\Sigma(N) + \beta
\Delta V_\Sigma(N)^2\right\}\tilde I(N)\ ,
\end{equation}
where $\alpha$ and $\beta$ are two real parameters to be suitably
chosen. Although both $\tilde I$ and $\tilde I'$ diverge in the
continuum limit, their difference vanishes. According to our
considerations, the fluctuations of $\Delta V_\Sigma$ are expected to
mimic the ones of $\tilde I$, and an appropriate and unique choice of
$\alpha$ and $\beta$ will provide a partial cancellation of the
irregularities observed in the continuum approach of $\tilde I'$, and
consequently $\tilde d_1$. The search of optimal values can be simply
performed by hand, as far as only two parameters have to be tuned (in
addition, $\beta$ multiplies a subdominant contribution). We find, in
particular, $(\alpha,\beta)=(0.0838,-0.45)$. A plot of $\tilde d_1$,
regularised according to this choice, is represented by the solid
curve in Figure~4. A fit of the smoothed data against (\ref{converg})
allows to extract the value of $d_1$ with much higher precision than
the previous determinations. The procedure can be extended to all the
other lattice integrals which contribute to the matching between the
continuum and the lattice heavy-light operators. Their explicit
expressions, reported in \cite{Flynn:1990qz,Gimenez:1998mw}, will not be reproduced here, but a
list of more accurate values, obtained with the method explained
above, is given in Table~\ref{tab:latint}.
\vskip 0.4cm
\TABLE[!h]{
  \centering
  \begin{tabular}{|c|c|c|c|c|c|}
    \hline
    $d_1$ & $J_1$ & $f$ & $v$ & $c$ & $w$ \\[1.0ex]
    \hline
    $5.4636(6)$ & $-4.8540(6)$ & $13.3503(6)$ & $-6.9230(24)$ &
    $4.5259(27)$ & $-1.20538(1)$ \\[1.0ex]
    \hline
  \end{tabular}
  \caption{Some basic infinite lattice integrals, which are
  needed to compute the matching coefficients $\cX^{(1)}_{{\rm
  ref,lat}}$. Their values have been determined according to the
  regularisation method explained in this Appendix.
  \label{tab:latint}
  }
}
\vskip 0.3cm
The finite lattice constants that contribute to the matching coefficients
$\cX^{(1)}_{{\rm ref,lat}}$  are expressed in terms
of those lattice integrals by a set of algebraic relations that
have been first published in \cite{Flynn:1990qz,Gimenez:1998mw}. Their values are reported in
Table~\ref{tab:infconst}. 

\vskip 0.5cm

\TABLE[!h]{
  \centering
  \begin{tabular}{|l|r||l|r|}
    \hline\hline
    \ $\ D_{LL}\ $  \   &\  $-41.248(8)$ \  &\  $\ \bar D_{LL}\ $ \   &\
    $2.5923(12)$\  \\ \hline
    \ $\ D^S_{LL}\ $\   &\  $-30.879(8)$ \  &\  $\ \bar D_{RL}\ $ \   &\
    $2.4489(11)$\  \\ \hline
    \ $\ D_{LR}\ $  \   &\  $-37.843(6)$ \  &\  $\ \bar D_{RR}\ $ \   &\
    $0.40179(3)\ $\  \\ \hline
    \ $\ D^S_{LR}\ $\   &\  $-37.843(6)$ \  &\  $\ \bar D^S_{RL}\ $\  &\
    $9.796(4)$ \   \\ \hline
    \ $\ D_{RR}\ $  \   &\  $-1.60717(1)$\  &  & \\
    \hline\hline
  \end{tabular}
  \caption{Values of the combinations of lattice integrals entering matching coefficients.}
  \label{tab:infconst}
}
The improvement integrals $f^I$, $v^I$ and $w^I$, introduced in
eq.~(\ref{matchingsw}), do not involve a factor of $\theta(1-k^2)$. Hence, their
numerical value can be computed with good precision through the blocking
procedure described in \cite{Luscher:1985wf}. We obtain:
\begin{align}
f^I & = -3.6461(2)\ ,  \nonumber \\
v^I & = -6.7185(2)\ , \nonumber \\
w^I & = 0.82130(2)\ .
\end{align}


%% file: appendC.tex
\section{Tables.}
\label{app:C}

In this appendix we list our results for the coefficients $r^+_{k,0}$,
as well as for the NLO anomalous dimension, obtained for different
discretisations and renormalisation schemes. We also take the
opportunity to describe
the blocking procedure \cite{Luscher:1985wf} applied to determine the
coefficients $r^+_{k,0}$ (see also ref. \cite{Palombi:2002gw} for
another practical application in the context of the SF).

Here we apply this method at the level of the first two blocking
steps, in order to eliminate the $\Oa$ cutoff effects in the
data. Going beyond this level yields no benefit with double precision
arithmetics, because of the numerical rounding that arises when
subsequent cancellations of the signal are performed. Starting from
the filtered data, we first check that the logarithmically divergent
term in the one-loop renormalisation constant has the correct
coefficient, which we indeed obtain with an average precision
\begin{align}
s_{k,0}^+ / \gamma_k'^{+;(0)} = 1.000(1)\ ,
\end{align}
cf. eq.~(\ref{eq:r0}).

After having checked the form of the divergence, we remove it from the
filtered data by subtracting explicitly a term
$\gamma_k'^{+;(0)}\ln(L/a)$. We then extract the finite part of the
renormalisation constant $r^+_{k,0}$ according to the following
procedure. The data are fitted with two different {\it ans\"atze},
viz.
\begin{align}
\cZ^{(n)}(L/a) = A + \sum_{\nu=2}^{n+1}\left(\dfrac{a}{L}\right)^\nu
  \left\{B_\nu + C_\nu\ln(L/a)\right\}\ , \qquad n = 1, 2.
\label{fit_model}
\end{align}
Rounding errors are modelled as suggested in~\cite{Bode:1999sm}. We
then fit in several intervals in $L/a$, always starting at the largest
available value. Next we study the $\chi^2$ per degree of freedom for
each fit {\it ansatz} as a function of the number of values of $L/a$
included, and find the minimum within the stability interval of the
fit, thus obtaining two estimates $[r^+_{k,0}]_{\rm 3p}$ and
$[r^+_{k,0}]_{\rm 5p}$ for $r^+_{k,0}$. We take $[r^+_{k,0}]_{\rm 3p}$
as our best estimate. Then we consider a number of estimates for the
uncertainty on $r^+_{k,0}$, namely: the fit errors on
$[r^+_{k,0}]_{\rm 3p}$ and $[r^+_{k,0}]_{\rm 5p}$; the difference
$|[r^+_{k,0}]_{\rm 3p}-[r^+_{k,0}]_{\rm 5p}|$; and the fluctuation of
$A$ within the stability interval of each fit. The final uncertainty
is taken to be the largest of all of them.

Our final results for the finite coefficients of the renormalisation
constants $r_{k,0}^+$ are reported in Tables
\ref{tab:constants1}--\ref{tab:constants6} below. In the determination
of the NLO anomalous dimensions, we have derived their uncertainties
by combining in quadrature the errors of $r_{k,0}^+$ and of the
matching coefficients in eq.~(\ref{matchRI}),
cf. Table~\ref{tab:infconst}.

\newpage

\newpage

\TABLE[!h]{
  \centering
  \begin{tabular}{cccccccc}
    \\\hline\hline\\[-2.0ex]
    action & $\theta$ & $s$ & $\alpha$ & $r^+_{1,0}$ &
    $r^+_{2,0}$ & $r^+_{3,0}$ & $r^+_{4,0}$ \\[0.5ex]
    \hline\\[-1.0ex]
    W--EH  & 0.0 & 1  & 0.0 & -0.09022(12) & -0.09029(4)  & -0.13884(12) & -0.12270(4)  \\
    W--EH  & 0.0 & 2  & 0.0 & -0.09022(12) & -0.09029(4)  & -0.13884(12) & -0.12270(4)  \\
    W--EH  & 0.0 & 3  & 0.0 & -0.09022(12) & -0.09029(4)  & -0.13884(12) & -0.12270(4)  \\
    W--EH  & 0.0 & 4  & 0.0 & -0.09022(12) & -0.09029(4)  & -0.13884(12) & -0.12270(4)  \\
    W--EH  & 0.0 & 5  & 0.0 & -0.09022(12) & -0.09029(4)  & -0.13884(12) & -0.12270(4)  \\[1.0ex]
    W--EH  & 0.0 & 1  & 0.5 & -0.09022(12) & -0.09029(4)  & -0.13884(12) & -0.12270(4)  \\
    W--EH  & 0.0 & 2  & 0.5 & -0.09022(12) & -0.09029(4)  & -0.13884(12) & -0.12270(4)  \\
    W--EH  & 0.0 & 3  & 0.5 & -0.09022(12) & -0.09029(4)  & -0.13884(12) & -0.12270(4)  \\
    W--EH  & 0.0 & 4  & 0.5 & -0.09022(12) & -0.09029(4)  & -0.13884(12) & -0.12270(4)  \\
    W--EH  & 0.0 & 5  & 0.5 & -0.09022(12) & -0.09029(4)  & -0.13884(12) & -0.12270(4)  \\[1.0ex]
    \hline\hline
  \end{tabular}
  \caption{
    Finite parts of the renormalisation constants in the primed operator
    basis. Light quarks are regularised with the unimproved Wilson
    action (W); heavy quarks with the Eichten-Hill one (EH). Here $\theta
    = 0.0$. Boundary conditions $s$ are enumerated according to (\ref{source3_1})-(\ref{source3_5}).
  }
  \label{tab:constants1}
}
\vskip 1.0cm
\TABLE[!h]{
  \centering
  \begin{tabular}{cccccccc}
    \\\hline\hline\\[-2.0ex]
    action & $\theta$ & $s$ & $\alpha$ & $r^+_{1,0}$ &
    $r^+_{2,0}$ & $r^+_{3,0}$ & $r^+_{4,0}$ \\[0.5ex]
    \hline\\[-1.0ex]
    SW--EH  & 0.0 & 1  & 0.0 & -0.05219(12)  & -0.05005(13)  & -0.1046(13)  & -0.08498(15) \\
    SW--EH  & 0.0 & 2  & 0.0 & -0.05219(12)  & -0.05005(13)  & -0.1046(13)  & -0.08498(15) \\
    SW--EH  & 0.0 & 3  & 0.0 & -0.05219(12)  & -0.05005(13)  & -0.1046(13)  & -0.08498(15) \\
    SW--EH  & 0.0 & 4  & 0.0 & -0.05219(12)  & -0.05005(13)  & -0.1046(13)  & -0.08498(15) \\
    SW--EH  & 0.0 & 5  & 0.0 & -0.05219(12)  & -0.05005(13)  & -0.1046(13)  & -0.08498(15) \\[1.0ex]
    SW--EH  & 0.0 & 1  & 0.5 & -0.05219(12)  & -0.05005(13)  & -0.1046(13)  & -0.08498(15) \\
    SW--EH  & 0.0 & 2  & 0.5 & -0.05219(12)  & -0.05005(13)  & -0.1046(13)  & -0.08498(15) \\
    SW--EH  & 0.0 & 3  & 0.5 & -0.05219(12)  & -0.05005(13)  & -0.1046(13)  & -0.08498(15) \\
    SW--EH  & 0.0 & 4  & 0.5 & -0.05219(12)  & -0.05005(13)  & -0.1046(13)  & -0.08498(15) \\
    SW--EH  & 0.0 & 5  & 0.5 & -0.05219(12)  & -0.05005(13)  & -0.1046(13)  & -0.08498(15) \\[1.0ex]
    \hline\hline
  \end{tabular}
  \caption{
    Finite parts of the renormalisation constants in the primed operator
    basis. Light quarks are regularised with the improved Wilson
    action (SW); heavy quarks with the Eichten-Hill one (EH). Here $\theta
    = 0.0$. Boundary conditions $s$ are enumerated according to
    (\ref{source3_1})-(\ref{source3_5}).
  }
  \label{tab:constants2}
}

\newpage

\TABLE[!h]{
  \centering
  \begin{tabular}{cccccccc}
    \\\hline\hline\\[-2.0ex]
    action & $\theta$ & $s$ & $\alpha$ & $r^+_{1,0}$ &
    $r^+_{2,0}$ & $r^+_{3,0}$ & $r^+_{4,0}$ \\[0.5ex]
    \hline\\[-1.0ex]
    W--EH  & 0.5 & 1  & 0.0 & -0.22302(8)  & -0.10054(7)  & -0.15083(5)  & -0.14615(10)  \\
    W--EH  & 0.5 & 2  & 0.0 & -0.21290(8)  & -0.09671(7)  & -0.14767(5)  & -0.14187(10)  \\
    W--EH  & 0.5 & 3  & 0.0 & -0.21290(8)  & -0.09566(7)  & -0.14767(5)  & -0.13734(10)  \\
    W--EH  & 0.5 & 4  & 0.0 & -0.22302(8)  & -0.10054(7)  & -0.15083(5)  & -0.14615(10)  \\
    W--EH  & 0.5 & 5  & 0.0 & -0.21290(8)  & -0.09356(7)  & -0.14767(5)  & -0.12828(10)  \\[1.0ex]
    W--EH  & 0.5 & 1  & 0.5 & -0.22723(8)  & -0.10475(5)  & -0.15504(10) & -0.15036(6)  \\
    W--EH  & 0.5 & 2  & 0.5 & -0.21711(8)  & -0.10092(5)  & -0.15188(10) & -0.14608(6)  \\
    W--EH  & 0.5 & 3  & 0.5 & -0.21711(8)  & -0.09987(5)  & -0.15188(10) & -0.14155(6)  \\
    W--EH  & 0.5 & 4  & 0.5 & -0.22723(8)  & -0.10475(5)  & -0.15504(10) & -0.15036(6)  \\
    W--EH  & 0.5 & 5  & 0.5 & -0.21711(8)  & -0.09777(5)  & -0.15188(10) & -0.13249(6)  \\[1.0ex]
    \hline\hline
  \end{tabular}
  \caption{
    Finite parts of the renormalisation constants in the primed operator
    basis. Light quarks are regularised with the unimproved Wilson
    action (W); heavy quarks with the Eichten-Hill one (EH). Here $\theta
    = 0.5$. Boundary conditions $s$ are enumerated according to
    (\ref{source3_1})-(\ref{source3_5}).
  }
  \label{tab:constants3}
}
\vskip 1.0cm
\TABLE[!h]{
  \centering
  \begin{tabular}{cccccccc}
    \\\hline\hline\\[-2.0ex]
    action & $\theta$ & $s$ & $\alpha$ & $r^+_{1,0}$ &
    $r^+_{2,0}$ & $r^+_{3,0}$ & $r^+_{4,0}$ \\[0.5ex]
    \hline\\[-1.0ex]
    SW--EH  & 0.5 & 1  & 0.0 & -0.1850(4)  & -0.06030(5)   & -0.11658(15)  & -0.10843(11)  \\
    SW--EH  & 0.5 & 2  & 0.0 & -0.1749(4)  & -0.05648(5)   & -0.11342(15)  & -0.10416(11)  \\
    SW--EH  & 0.5 & 3  & 0.0 & -0.1749(4)  & -0.05543(5)   & -0.11342(15)  & -0.09963(11)  \\
    SW--EH  & 0.5 & 4  & 0.0 & -0.1850(4)  & -0.06031(5)   & -0.11658(15)  & -0.10843(11)  \\
    SW--EH  & 0.5 & 5  & 0.0 & -0.1749(4)  & -0.05333(5)   & -0.11342(15)  & -0.09057(11)  \\[1.0ex]
    SW--EH  & 0.5 & 1  & 0.5 & -0.1892(4)   & -0.06451(7)   & -0.12078(15) & -0.11264(13) \\
    SW--EH  & 0.5 & 2  & 0.5 & -0.1791(4)   & -0.06069(7)   & -0.11763(15) & -0.10836(13) \\
    SW--EH  & 0.5 & 3  & 0.5 & -0.1791(4)   & -0.05964(7)   & -0.11763(15) & -0.10383(13) \\
    SW--EH  & 0.5 & 4  & 0.5 & -0.1892(4)   & -0.06451(7)   & -0.12078(15) & -0.11264(13) \\
    SW--EH  & 0.5 & 5  & 0.5 & -0.1791(4)   & -0.05754(7)   & -0.11763(15) & -0.09478(12) \\[1.0ex]
    \hline\hline
  \end{tabular}
  \caption{
    Finite parts of the renormalisation constants in the primed operator
    basis. Light quarks are regularised with the improved Wilson
    action (SW); heavy quarks with the Eichten-Hill one (EH). Here $\theta
    = 0.5$. Boundary conditions $s$ are enumerated according to
    (\ref{source3_1})-(\ref{source3_5}).
  }
  \label{tab:constants4}
}

\newpage

\TABLE[!h]{
  \centering
  \begin{tabular}{cccccccc}
    \\\hline\hline\\[-2.0ex]
    action & $\theta$ & $s$ & $\alpha$ & $r^+_{1,0}$ &
    $r^+_{2,0}$ & $r^+_{3,0}$ & $r^+_{4,0}$ \\[0.5ex]
    \hline\\[-1.0ex]
    W--EH  & 1.0 & 1  & 0.0 & -0.3596(10)  & -0.10601(14) & -0.16560(14) & -0.15650(23) \\
    W--EH  & 1.0 & 2  & 0.0 & -0.3421(10)  & -0.10173(14) & -0.16161(14) & -0.15057(24) \\
    W--EH  & 1.0 & 3  & 0.0 & -0.3421(10)  & -0.09975(14) & -0.16161(14) & -0.14540(24) \\
    W--EH  & 1.0 & 4  & 0.0 & -0.3596(10)  & -0.10601(14) & -0.16560(14) & -0.15650(23) \\
    W--EH  & 1.0 & 5  & 0.0 & -0.3421(10)  & -0.09579(14) & -0.16161(14) & -0.13509(24)  \\[1.0ex]
    W--EH  & 1.0 & 1  & 0.5 & -0.3649(10)  & -0.11133(14) & -0.17093(14) & -0.16183(23) \\
    W--EH  & 1.0 & 2  & 0.5 & -0.3474(10)  & -0.10706(14) & -0.16693(14) & -0.15589(23) \\
    W--EH  & 1.0 & 3  & 0.5 & -0.3474(10)  & -0.10507(14) & -0.16693(14) & -0.15073(23) \\
    W--EH  & 1.0 & 4  & 0.5 & -0.3649(10)  & -0.11133(14) & -0.17092(14) & -0.16183(23) \\
    W--EH  & 1.0 & 5  & 0.5 & -0.3474(10)  & -0.10110(14) & -0.16694(14) & -0.14041(24)  \\[1.0ex]
    \hline\hline
  \end{tabular}
  \caption{
    Finite parts of the renormalisation constants in the primed operator
    basis. Light quarks are regularised with the unimproved Wilson
    action (W); heavy quarks with the Eichten-Hill one (EH). Here $\theta
    = 1.0$. Boundary conditions $s$ are enumerated according to
    (\ref{source3_1})-(\ref{source3_5}).
  }
  \label{tab:constants5}
}
\vskip 1.0cm
\TABLE[!h]{
  \centering
  \begin{tabular}{cccccccc}
    \\\hline\hline\\[-2.0ex]
    action & $\theta$ & $s$ & $\alpha$ & $r^+_{1,0}$ &
    $r^+_{2,0}$ & $r^+_{3,0}$ & $r^+_{4,0}$ \\[0.5ex]
    \hline\\[-1.0ex]
    SW--EH  & 1.0 & 1  & 0.0 & -0.3216(4)  & -0.06578(19) & -0.13135(23) & -0.1188(4)  \\
    SW--EH  & 1.0 & 2  & 0.0 & -0.3041(4)  & -0.06151(19) & -0.12736(23) & -0.1129(4)  \\
    SW--EH  & 1.0 & 3  & 0.0 & -0.3041(4)  & -0.05953(19) & -0.12736(23) & -0.1077(4)  \\
    SW--EH  & 1.0 & 4  & 0.0 & -0.3216(4)  & -0.06578(19) & -0.13135(23) & -0.1188(4)  \\
    SW--EH  & 1.0 & 5  & 0.0 & -0.3041(4)  & -0.05557(19) & -0.12736(23) & -0.0974(4)  \\[1.0ex]
    SW--EH  & 1.0 & 1  & 0.5 & -0.3269(4)  & -0.07109(19) & -0.13666(23) & -0.1241(4)  \\
    SW--EH  & 1.0 & 2  & 0.5 & -0.3094(4)  & -0.06683(19) & -0.13268(23) & -0.1182(4)  \\
    SW--EH  & 1.0 & 3  & 0.5 & -0.3094(4)  & -0.06485(19) & -0.13268(23) & -0.1130(4)  \\
    SW--EH  & 1.0 & 4  & 0.5 & -0.3269(4)  & -0.07109(19) & -0.13666(23) & -0.1241(4)  \\
    SW--EH  & 1.0 & 5  & 0.5 & -0.3094(4)  & -0.06089(19) & -0.13268(23) & -0.1027(4)  \\[1.0ex]
    \hline\hline
  \end{tabular}
  \caption{
    Finite parts of the renormalisation constants in the primed operator
    basis. Light quarks are regularised with the improved Wilson
    action (SW); heavy quarks with the Eichten-Hill one (EH). Here $\theta
    = 1.0$. Boundary conditions $s$ are enumerated according to
    (\ref{source3_1})-(\ref{source3_5}).
  }
  \label{tab:constants6}
}

\newpage

\TABLE[!h]{
  \centering
  \begin{tabular}{ccccc}
    \\\hline\hline\\[-2.0ex]
    $\theta$ & $s$ & $\alpha$ & $\gamma'^{+;(1)}_{1;\rm SF}/\gamma'^{+;(0)}_1$ &
    $\gamma'^{+;(1)}_{2;\rm SF}/\gamma'^{+;(0)}_2$ \\[0.5ex]
    \hline\\[-1.0ex]
    0.0 &  1  & 0.0 & $-0.2087(4) +0.02294(2)\Nf$ & $-0.5535(5) + 0.04876(3)\Nf$ \\
    0.0 &  2  & 0.0 & $-0.2087(4) +0.02294(2)\Nf$ & $-0.5535(5) + 0.04876(3)\Nf$ \\
    0.0 &  3  & 0.0 & $-0.2087(4) +0.02294(2)\Nf$ & $-0.5535(5) + 0.04876(3)\Nf$ \\
    0.0 &  4  & 0.0 & $-0.2087(4) +0.02294(2)\Nf$ & $-0.5535(5) + 0.04876(3)\Nf$ \\
    0.0 &  5  & 0.0 & $-0.2087(4) +0.02294(2)\Nf$ & $-0.5535(5) + 0.04876(3)\Nf$ \\
    \\[-1.0ex]\hline\\[-2.0ex]
    $\theta$ & $s$ & $\alpha$ & $\gamma'^{+;(1)}_{3;\rm SF}/\gamma'^{+;(0)}_3$ &
    $\gamma'^{+;(1)}_{4;\rm SF}/\gamma'^{+;(0)}_4$ \\[0.5ex]
    \hline\\[-1.0ex]
    0.0 &  1  & 0.0 & $-0.0590(3) + 0.01371(2)\Nf$ & $-0.1654(3) + 0.02478(2)\Nf$ \\
    0.0 &  2  & 0.0 & $-0.0590(3) + 0.01371(2)\Nf$ & $-0.1654(3) + 0.02478(2)\Nf$ \\
    0.0 &  3  & 0.0 & $-0.0590(3) + 0.01371(2)\Nf$ & $-0.1654(3) + 0.02478(2)\Nf$ \\
    0.0 &  4  & 0.0 & $-0.0590(3) + 0.01371(2)\Nf$ & $-0.1654(3) + 0.02478(2)\Nf$ \\
    0.0 &  5  & 0.0 & $-0.0590(3) + 0.01371(2)\Nf$ & $-0.1654(3) + 0.02478(2)\Nf$ \\[1.0ex]
    \hline\hline
  \end{tabular}
  \caption{The two loop anomalous dimensions of the diagonal basis in units of the corresponding 
    universal one-loop coefficients. Here $\theta = 0.0$ and $\alpha = 0.0$.}
  \label{tab:NLO1}
}
\TABLE[!h]{
  \centering
  \begin{tabular}{ccccc}
    \\\hline\hline\\[-2.0ex]
    $\theta$ & $s$ & $\alpha$ & $\gamma'^{+;(1)}_{1;\rm SF}/\gamma'^{+;(0)}_1$ &
    $\gamma'^{+;(1)}_{2;\rm SF}/\gamma'^{+;(0)}_2$ \\[0.5ex]
    \hline\\[-1.0ex]
    0.0 &  1  & 0.5 & $-0.2087(4) +0.02294(2)\Nf$ & $-0.5535(5) + 0.04876(3)\Nf$ \\
    0.0 &  2  & 0.5 & $-0.2087(4) +0.02294(2)\Nf$ & $-0.5535(5) + 0.04876(3)\Nf$ \\
    0.0 &  3  & 0.5 & $-0.2087(4) +0.02294(2)\Nf$ & $-0.5535(5) + 0.04876(3)\Nf$ \\
    0.0 &  4  & 0.5 & $-0.2087(4) +0.02294(2)\Nf$ & $-0.5535(5) + 0.04876(3)\Nf$ \\
    0.0 &  5  & 0.5 & $-0.2087(4) +0.02294(2)\Nf$ & $-0.5535(5) + 0.04876(3)\Nf$ \\
    \\[-1.0ex]\hline\\[-2.0ex]
    $\theta$ & $s$ & $\alpha$ & $\gamma'^{+;(1)}_{3;\rm SF}/\gamma'^{+;(0)}_3$ &
    $\gamma'^{+;(1)}_{4;\rm SF}/\gamma'^{+;(0)}_4$ \\[0.5ex]
    \hline\\[-1.0ex]
    0.0 &  1  & 0.5 & $-0.0590(3) + 0.01371(2)\Nf$ & $-0.1654(3) + 0.02478(2)\Nf$ \\
    0.0 &  2  & 0.5 & $-0.0590(3) + 0.01371(2)\Nf$ & $-0.1654(3) + 0.02478(2)\Nf$ \\
    0.0 &  3  & 0.5 & $-0.0590(3) + 0.01371(2)\Nf$ & $-0.1654(3) + 0.02478(2)\Nf$ \\
    0.0 &  4  & 0.5 & $-0.0590(3) + 0.01371(2)\Nf$ & $-0.1654(3) + 0.02478(2)\Nf$ \\
    0.0 &  5  & 0.5 & $-0.0590(3) + 0.01371(2)\Nf$ & $-0.1654(3) + 0.02478(2)\Nf$ \\[1.0ex]
    \hline\hline
  \end{tabular}
  \caption{The two loop anomalous dimensions of the diagonal basis in units of the corresponding
    universal one-loop coefficients. Here $\theta = 0.0$ and $\alpha = 0.5$.}
  \label{tab:NLO2}
}

\newpage

\TABLE[!h]{
  \centering
  \begin{tabular}{ccccc}
    \\\hline\hline\\[-2.0ex]
    $\theta$ & $s$ & $\alpha$ & $\gamma'^{+;(1)}_{1;\rm SF}/\gamma'^{+;(0)}_1$ &
    $\gamma'^{+;(1)}_{2;\rm SF}/\gamma'^{+;(0)}_2$ \\[0.5ex]
    \hline\\[-1.0ex]
    0.5 &  1  & 0.0 & $0.1566(4) + 0.00080(2)\Nf$ & $-0.4690(5) + 0.04364(3)\Nf$ \\
    0.5 &  2  & 0.0 & $0.1287(4) + 0.00249(2)\Nf$ & $-0.5006(5) + 0.04555(3)\Nf$ \\
    0.5 &  3  & 0.0 & $0.1287(4) + 0.00249(2)\Nf$ & $-0.5092(5) + 0.04608(3)\Nf$ \\
    0.5 &  4  & 0.0 & $0.1565(4) + 0.00080(2)\Nf$ & $-0.4690(5) + 0.04364(3)\Nf$ \\
    0.5 &  5  & 0.0 & $0.1287(4) + 0.00249(2)\Nf$ & $-0.5265(5) + 0.04713(3)\Nf$ \\
    \\[-1.0ex]\hline\\[-2.0ex]
    $\theta$ & $s$ & $\alpha$ & $\gamma'^{+;(1)}_{3;\rm SF}/\gamma'^{+;(0)}_3$ &
    $\gamma'^{+;(1)}_{4;\rm SF}/\gamma'^{+;(0)}_4$ \\[0.5ex]
    \hline\\[-1.0ex]
    0.5 &  1  & 0.0 & $-0.0327(3)  + 0.01211(2)\Nf$ & $-0.0364(3) + 0.01696(2) \Nf$ \\
    0.5 &  2  & 0.0 & $-0.0396(3)  + 0.01254(2)\Nf$ & $-0.0600(3) + 0.01839(2) \Nf$ \\
    0.5 &  3  & 0.0 & $-0.0396(3)  + 0.01254(2)\Nf$ & $-0.0849(3) + 0.01990(2) \Nf$ \\
    0.5 &  4  & 0.0 & $-0.0327(3)  + 0.01211(2)\Nf$ & $-0.0364(3) + 0.01696(2) \Nf$ \\
    0.5 &  5  & 0.0 & $-0.0396(3)  + 0.01254(2)\Nf$ & $-0.1347(3) + 0.02292(2) \Nf$ \\[1.0ex]
    \hline\hline
  \end{tabular}
  \caption{The two loop anomalous dimensions of the diagonal basis in units of the corresponding 
    universal one-loop coefficients. Here $\theta = 0.5$ and $\alpha = 0.0$.}
  \label{tab:NLO3}
}
\TABLE[!h]{
  \centering
  \begin{tabular}{ccccc}
    \\\hline\hline\\[-2.0ex]
    $\theta$ & $s$ & $\alpha$ & $\gamma'^{+;(1)}_{1;\rm SF}/\gamma'^{+;(0)}_1$ &
    $\gamma'^{+;(1)}_{2;\rm SF}/\gamma'^{+;(0)}_2$ \\[0.5ex]
    \hline\\[-1.0ex]
    0.5 &  1  & 0.5 & $0.1681(3) + 0.00010(2) \Nf$ & $-0.4342(6) + 0.04153(3) \Nf$ \\
    0.5 &  2  & 0.5 & $0.1403(3) + 0.00179(2) \Nf$ & $-0.4658(6) + 0.04344(3) \Nf$ \\
    0.5 &  3  & 0.5 & $0.1403(3) + 0.00179(2) \Nf$ & $-0.4745(6) + 0.04397(3) \Nf$ \\
    0.5 &  4  & 0.5 & $0.1681(3) + 0.00010(2) \Nf$ & $-0.4342(6) + 0.04153(3) \Nf$ \\
    0.5 &  5  & 0.5 & $0.1403(3) + 0.00179(2) \Nf$ & $-0.4918(6) + 0.04502(3) \Nf$ \\
    \\[-1.0ex]\hline\\[-2.0ex]
    $\theta$ & $s$ & $\alpha$ & $\gamma'^{+;(1)}_{3;\rm SF}/\gamma'^{+;(0)}_3$ &
    $\gamma'^{+;(1)}_{4;\rm SF}/\gamma'^{+;(0)}_4$ \\[0.5ex]
    \hline\\[-1.0ex]
    0.5 &  1  & 0.5 & $-0.0234(2) + 0.01155(2) \Nf$ & $-0.0133(3) + 0.01556(2) \Nf$ \\
    0.5 &  2  & 0.5 & $-0.0304(2) + 0.01197(2) \Nf$ & $-0.0368(3) + 0.01699(2) \Nf$ \\
    0.5 &  3  & 0.5 & $-0.0304(2) + 0.01197(2) \Nf$ & $-0.0617(3) + 0.01850(2) \Nf$ \\
    0.5 &  4  & 0.5 & $-0.0234(2) + 0.01155(2) \Nf$ & $-0.0133(3) + 0.01556(2) \Nf$ \\
    0.5 &  5  & 0.5 & $-0.0304(2) + 0.01197(2) \Nf$ & $-0.1116(3) + 0.02152(2) \Nf$ \\[1.0ex]
    \hline\hline
  \end{tabular}
  \caption{The two loop anomalous dimensions of the diagonal basis in units of the corresponding
    universal one-loop coefficients. Here $\theta = 0.5$ and $\alpha = 0.5$.}
  \label{tab:NLO4}
}

\newpage

\vfill

\TABLE[!h]{
  \centering
  \begin{tabular}{ccccc}
    \\\hline\hline\\[-2.0ex]
    $\theta$ & $s$ & $\alpha$ & $\gamma'^{+;(1)}_{1;\rm SF}/\gamma'^{+;(0)}_1$ &
    $\gamma'^{+;(1)}_{2;\rm SF}/\gamma'^{+;(0)}_2$ \\[0.5ex]
    \hline\\[-1.0ex]
    1.0 &  1  & 0.0 & $0.5321(3) - 0.02196(2) \Nf$ & $-0.424(1) + 0.04090(8) \Nf$ \\
    1.0 &  2  & 0.0 & $0.4840(3) - 0.01904(2) \Nf$ & $-0.459(1) + 0.04304(8) \Nf$ \\
    1.0 &  3  & 0.0 & $0.4840(3) - 0.01904(2) \Nf$ & $-0.475(1) + 0.04403(8) \Nf$ \\
    1.0 &  4  & 0.0 & $0.5321(3) - 0.02196(2) \Nf$ & $-0.424(1) + 0.04090(8) \Nf$ \\
    1.0 &  5  & 0.0 & $0.4840(3) - 0.01904(2) \Nf$ & $-0.508(1) + 0.04601(8) \Nf$ \\
    \\[-1.0ex]\hline\\[-2.0ex]
    $\theta$ & $s$ & $\alpha$ & $\gamma'^{+;(1)}_{3;\rm SF}/\gamma'^{+;(0)}_3$ &
    $\gamma'^{+;(1)}_{4;\rm SF}/\gamma'^{+;(0)}_4$ \\[0.5ex]
    \hline\\[-1.0ex]
    1.0 &  1  & 0.0 & $-0.0002(3) + 0.01014(2) \Nf$ & $\hskip 0.30cm 0.020(1) + 0.01351(8) \Nf$ \\
    1.0 &  2  & 0.0 & $-0.0090(3) + 0.01068(2) \Nf$ & $-0.012(1) + 0.01549(8) \Nf$ \\
    1.0 &  3  & 0.0 & $-0.0090(3) + 0.01068(2) \Nf$ & $-0.041(1) + 0.01721(8) \Nf$ \\
    1.0 &  4  & 0.0 & $-0.0002(3) + 0.01014(2) \Nf$ & $\hskip 0.30cm 0.020(1) + 0.01351(8) \Nf$ \\
    1.0 &  5  & 0.0 & $-0.0090(3) + 0.01068(2) \Nf$ & $-0.097(1) + 0.02065(8) \Nf$ \\[1.0ex]
    \hline\hline
  \end{tabular}
  \caption{The two loop anomalous dimensions of the diagonal basis in units of the corresponding 
    universal one-loop coefficients. Here $\theta = 1.0$ and $\alpha = 0.0$.}
  \label{tab:NLO5}
}
\TABLE[!h]{
  \centering
  \begin{tabular}{ccccc}
    \\\hline\hline\\[-2.0ex]
    $\theta$ & $s$ & $\alpha$ & $\gamma'^{+;(1)}_{1;\rm SF}/\gamma'^{+;(0)}_1$ &
    $\gamma'^{+;(1)}_{2;\rm SF}/\gamma'^{+;(0)}_2$ \\[0.5ex]
    \hline\\[-1.0ex]
    1.0 &  1  & 0.5 & $0.5467(3) - 0.02284(2) \Nf$ & $-0.380(1) + 0.03824(8) \Nf$ \\
    1.0 &  2  & 0.5 & $0.4986(3) - 0.01993(2) \Nf$ & $-0.415(1) + 0.04038(8) \Nf$ \\
    1.0 &  3  & 0.5 & $0.4986(3) - 0.01993(2) \Nf$ & $-0.432(1) + 0.04137(8) \Nf$ \\
    1.0 &  4  & 0.5 & $0.5467(3) - 0.02284(2) \Nf$ & $-0.380(1) + 0.03824(8) \Nf$ \\
    1.0 &  5  & 0.5 & $0.4986(3) - 0.01993(2) \Nf$ & $-0.464(1) + 0.04336(8) \Nf$ \\
    \\[-1.0ex]\hline\\[-2.0ex]
    $\theta$ & $s$ & $\alpha$ & $\gamma'^{+;(1)}_{3;\rm SF}/\gamma'^{+;(0)}_3$ &
    $\gamma'^{+;(1)}_{4;\rm SF}/\gamma'^{+;(0)}_4$ \\[0.5ex]
    \hline\\[-1.0ex]
    1.0 &  1  & 0.5 & $0.0116(3) + 0.00943(2) \Nf$ & $\hskip 0.30cm 0.050(1)  + 0.01174(8) \Nf$ \\
    1.0 &  2  & 0.5 & $0.0028(3) + 0.00997(2) \Nf$ & $\hskip 0.30cm 0.017(1)  + 0.01372(8) \Nf$ \\
    1.0 &  3  & 0.5 & $0.0028(3) + 0.00997(2) \Nf$ & $-0.011(1) + 0.01544(8) \Nf$ \\
    1.0 &  4  & 0.5 & $0.0115(3) + 0.00943(2) \Nf$ & $\hskip 0.30cm 0.050(1)  + 0.01174(8) \Nf$ \\
    1.0 &  5  & 0.5 & $0.0028(3) + 0.00997(2) \Nf$ & $-0.068(1) + 0.01888(8) \Nf$ \\[1.0ex]
    \hline\hline
  \end{tabular}
  \caption{The two loop anomalous dimensions of the diagonal basis in units of the corresponding
    universal one-loop coefficients. Here $\theta = 1.0$ and $\alpha = 0.5$.}
  \label{tab:NLO6}
}

\vfill\eject

%% file: biblio.tex
\vfill\eject

%% file: main.bbl
\begin{thebibliography}{99}

\bibitem{Abada:1991mt}
A.~Abada {\it et al.},
Nucl.\ Phys.\ B {\bf 376} (1992) 172.

\bibitem{Ewing:1995ih}
A.K.~Ewing {\it et al.}  [UKQCD Collaboration],
Phys.\ Rev.\ D {\bf 54} (1996) 3526
[arXiv:hep-lat/9508030].

\bibitem{Gimenez:1996sk}
V.~Gim\'enez and G.~Martinelli,
Phys.\ Lett.\ B {\bf 398} (1997) 135
[arXiv:hep-lat/9610024].

\bibitem{Christensen:1996sj}
J.C.~Christensen, T.~Draper and C.~McNeile,
Phys.\ Rev.\ D {\bf 56} (1997) 6993
[arXiv:hep-lat/9610026].

\bibitem{Bernard:1998dg}
C.W.~Bernard, T.~Blum and A.~Soni,
Phys.\ Rev.\ D {\bf 58} (1998) 014501
[arXiv:hep-lat/9801039].

\bibitem{Gimenez:1998mw}
V.~Gim\'enez and J.~Reyes,
Nucl.\ Phys.\ B {\bf 545} (1999) 576
[arXiv:hep-lat/9806023]; \\
J.~Reyes, ``C\'alculo de elementos de matriz d\'ebiles para hadrones
$B$ con la HQET en el ret\'{\i}culo'', Ph.~D.~Thesis, University of
Valencia, May 2001. 

\bibitem{Becirevic:2000nv}
D.~Be\'cirevi\'c, D.~Meloni, A.~Retico, V.~Gim\'enez, L.~Giusti,
V.~Lubicz and G.~Martinelli,
Nucl.\ Phys.\ B {\bf 618} (2001) 241
[arXiv:hep-lat/0002025].

\bibitem{Hashimoto:2000eh}
S.~Hashimoto, K.I.~Ishikawa, T.~Onogi, M.~Sakamoto, N.~Tsutsui and
N.~Yamada,
Phys.\ Rev.\ D {\bf 62} (2000) 114502
[arXiv:hep-lat/0004022].

\bibitem{Lellouch:2000tw}
L.~Lellouch and C.J.D.~Lin  [UKQCD Collaboration],
Phys.\ Rev.\ D {\bf 64} (2001) 094501
[arXiv:hep-ph/0011086].

\bibitem{Becirevic:2001xt}
D.~Be\'cirevi\'c, V.~Gim\'enez, G.~Martinelli, M.~Papinutto and J.~Reyes,
JHEP {\bf 0204} (2002) 025
[arXiv:hep-lat/0110091].

\bibitem{Aoki:2002bh}
S.~Aoki {\it et al.}  [JLQCD Collaboration],
Phys.\ Rev.\ D {\bf 67} (2003) 014506
[arXiv:hep-lat/0208038].

\bibitem{Aoki:2003xb}
S.~Aoki {\it et al.}  [JLQCD Collaboration],
Phys.\ Rev.\ Lett.\  {\bf 91} (2003) 212001
[arXiv:hep-ph/0307039].

\bibitem{Eichten:1987xu}
E.~Eichten,
Nucl.\ Phys.\ Proc.\ Suppl.\  {\bf 4} (1988) 170.

\bibitem{Heitger:2003nj}
J.~Heitger and R.~Sommer  [ALPHA Collaboration],
JHEP {\bf 0402} (2004) 022
[arXiv:hep-lat/0310035].

\bibitem{NRQCD}
B.A.~Thacker and G.P.~Lepage,
Phys.\ Rev.\ D {\bf 43} (1991) 196;
G.P.~Lepage, L.~Magnea, C.~Nakhleh, U.~Magnea and K.~Hornbostel,
Phys.\ Rev.\ D {\bf 46} (1992) 4052
[arXiv:hep-lat/9205007].

\bibitem{El-Khadra:1996mp}
A.X.~El-Khadra, A.S.~Kronfeld and P.B.~Mackenzie,
Phys.\ Rev.\ D {\bf 55} (1997) 3933
[arXiv:hep-lat/9604004].

\bibitem{Aoki:2001ra}
S.~Aoki, Y.~Kuramashi and S.~Tominaga,
Prog.\ Theor.\ Phys.\  {\bf 109} (2003) 383
[arXiv:hep-lat/0107009].

\bibitem{Guagnelli:2002jd}
M.~Guagnelli, F.~Palombi, R.~Petronzio and N.~Tantalo,
Phys.\ Lett.\ B {\bf 546} (2002) 237
[arXiv:hep-lat/0206023].

\bibitem{Donini:1999sf}
A.~Donini, V.~Gim\'enez, G.~Martinelli, M.~Talevi and A.~Vladikas,
Eur.\ Phys.\ J.\ C {\bf 10} (1999) 121
[arXiv:hep-lat/9902030].

\bibitem{Frezzotti:2000nk}
R.~Frezzotti, P.A.~Grassi, S.~Sint and P.~Weisz  [ALPHA
Collaboration],
JHEP {\bf 0108} (2001) 058
[arXiv:hep-lat/0101001].

\bibitem{Luscher:1992an}
M.~L\"uscher, R.~Narayanan, P.~Weisz and U.~Wolff,
Nucl.\ Phys.\ B {\bf 384} (1992) 168
[arXiv:hep-lat/9207009].

\bibitem{Guagnelli:2005zc}
M.~Guagnelli, J.~Heitger, C.~Pena, S.~Sint and A.~Vladikas  [ALPHA
Collaboration],
JHEP {\bf 0603} (2006) 088
[arXiv:hep-lat/0505002].

\bibitem{Palombi:2005zd}
F.~Palombi, C.~Pena and S.~Sint,
JHEP {\bf 0603} (2006) 089
[arXiv:hep-lat/0505003].

\bibitem{Becirevic:2003hd}
D.~Be\'cirevi\'c and J.~Reyes,
Nucl.\ Phys.\ Proc.\ Suppl.\  {\bf 129} (2004) 435
[arXiv:hep-lat/0309131].

\bibitem{Becirevic:2005sx}
D.~Be\'cirevi\'c, B.~Blossier, P.~Boucaud, J.P.~Leroy, A.~Le Yaouanc
and O.~P\`ene,
PoS {\bf LAT2005} (2005) 218
[arXiv:hep-lat/0509165].

\bibitem{Eichten:1989zv}
E.~Eichten and B.~Hill,
Phys.\ Lett.\ B {\bf 234} (1990) 511.

\bibitem{DellaMorte:2005yc}
M.~Della Morte, A.~Shindler and R.~Sommer,
JHEP {\bf 0508} (2005) 051
[arXiv:hep-lat/0506008].

\bibitem{Sheikholeslami:1985ij}
  B.~Sheikholeslami and R.~Wohlert,
  Nucl.\ Phys.\ B {\bf 259} (1985) 572.
  
\bibitem{Flynn:1990qz}
J.M.~Flynn, O.F.~Hern\'andez and B.R.~Hill,
Phys.\ Rev.\ D {\bf 43} (1991) 3709.

\bibitem{Bochicchio:1985xa}
M.~Bochicchio, L.~Maiani, G.~Martinelli, G.C.~Rossi and M.~Testa,
Nucl.\ Phys.\ B {\bf 262} (1985) 331.

\bibitem{alpha:su3}
M.~L\"uscher, R.~Sommer, P.~Weisz and U.~Wolff,
Nucl.\ Phys.\ B {\bf 413} (1994) 481
[arXiv:hep-lat/9309005].

\bibitem{mbar:pap1}
S.~Capitani, M.~L\"uscher, R.~Sommer and H.~Wittig  [ALPHA Collaboration],
Nucl.\ Phys.\ B {\bf 544} (1999) 669
[arXiv:hep-lat/9810063].

\bibitem{DellaMorte:2004bc}
M.~Della Morte, R.~Frezzotti, J.~Heitger, J.~Rolf, R.~Sommer and
U.~Wolff [ALPHA Collaboration],
Nucl.\ Phys.\ B {\bf 713} (2005) 378
[arXiv:hep-lat/0411025].

\bibitem{mbar:pap3}
J.~Garden, J.~Heitger, R.~Sommer and H.~Wittig  [ALPHA Collaboration],
Nucl.\ Phys.\ B {\bf 571} (2000) 237
[arXiv:hep-lat/9906013].

\bibitem{RolfSint_mc}
J.~Rolf and S.~Sint  [ALPHA Collaboration],
JHEP {\bf 0212} (2002) 007
[arXiv:hep-ph/0209255].

\bibitem{DellaMorte:2005kg}
M.~Della Morte, R.~Hoffmann, F.~Knechtli, J.~Rolf, R.~Sommer,
I.~Wetzorke and U.~Wolff [ALPHA Collaboration],
arXiv:hep-lat/0507035.

\bibitem{mbar:pap2}
M.~Guagnelli, J.~Heitger, R.~Sommer and H.~Wittig  [ALPHA Collaboration],
Nucl.\ Phys.\ B {\bf 560} (1999) 465
[arXiv:hep-lat/9903040].

\bibitem{juettner_fDs}
A.~J\"uttner and J.~Rolf  [ALPHA Collaboration],
Phys.\ Lett.\ B {\bf 560} (2003) 59
[arXiv:hep-lat/0302016].

\bibitem{deDivitiis:2003wy}
G.M.~de Divitiis, M.~Guagnelli, F.~Palombi, R.~Petronzio and
N.~Tantalo,
Nucl.\ Phys.\ B {\bf 672} (2003) 372
[arXiv:hep-lat/0307005].

\bibitem{Guagnelli:2004ga}
M.~Guagnelli, K.~Jansen, F.~Palombi, R.~Petronzio, A.~Shindler and
I.~Wetzorke [Zeuthen-Rome (ZeRo) Collaboration],
Eur.\ Phys.\ J.\ C {\bf 40} (2005) 69
[arXiv:hep-lat/0405027].

\bibitem{Guagnelli:2004ww}
M.~Guagnelli, K.~Jansen, F.~Palombi, R.~Petronzio, A.~Shindler and
I.~Wetzorke [Zeuthen-Rome (ZeRo) Collaboration],
Phys.\ Lett.\ B {\bf 597} (2004) 216
[arXiv:hep-lat/0403009].

\bibitem{Dimopoulos:2006dm}
  P.~Dimopoulos, J.~Heitger, F.~Palombi, C.~Pena, S.~Sint and A.~Vladikas [ALPHA Collaboration],
  arXiv:hep-ph/0601002.

\bibitem{Kurth:2000ki}
M.~Kurth and R.~Sommer  [ALPHA Collaboration],
Nucl.\ Phys.\ B {\bf 597} (2001) 488
[arXiv:hep-lat/0007002].

\bibitem{Heitger:2003xg}
  J.~Heitger, M.~Kurth and R.~Sommer  [ALPHA Collaboration],
  Nucl.\ Phys.\ B {\bf 669} (2003) 173
  [arXiv:hep-lat/0302019].

\bibitem{Luscher:1996sc}
M.~L\"uscher, S.~Sint, R.~Sommer and P.~Weisz,
Nucl.\ Phys.\ B {\bf 478} (1996) 365
[arXiv:hep-lat/9605038].

\bibitem{deDivitiis:1994yz}
G.~de Divitiis {\it et al.}  [ALPHA Collaboration],
Nucl.\ Phys.\ B {\bf 437} (1995) 447
[arXiv:hep-lat/9411017].

\bibitem{Guagnelli:2004za}
M.~Guagnelli, J.~Heitger, F.~Palombi, C.~Pena and A.~Vladikas
[ALPHA Collaboration],
JHEP {\bf 0405}, 001 (2004)
[arXiv:hep-lat/0402022].

\bibitem{Sint:1998iq}
S.~Sint and P.~Weisz  [ALPHA collaboration],
Nucl.\ Phys.\ B {\bf 545} (1999) 529
[arXiv:hep-lat/9808013].

\bibitem{Obeso:2005mc}
  E.~Obeso,
  PoS {\bf LAT2005} (2005) 234
  [arXiv:hep-lat/0509191].

\bibitem{Panagopoulos:2001fn}
H.~Panagopoulos and Y.~Proestos,
Phys.\ Rev.\ D {\bf 65} (2002) 014511
[arXiv:hep-lat/0108021].

\bibitem{Sint:1997dj}
S.~Sint and P.~Weisz,
Nucl.\ Phys.\ Proc.\ Suppl.\  {\bf 63}, 856 (1998)
[arXiv:hep-lat/9709096].

\bibitem{Luscher:1985wf}
M.~L\"uscher and P.~Weisz,
Nucl.\ Phys.\ B {\bf 266} (1986) 309.

\bibitem{Sint:1995ch}
S.~Sint and R.~Sommer,
Nucl.\ Phys.\ B {\bf 465} (1996) 71
[arXiv:hep-lat/9508012].

\bibitem{Pena:2004gb}
  C.~Pena, S.~Sint and A.~Vladikas,
  JHEP {\bf 0409} (2004) 069
  [arXiv:hep-lat/0405028].

\bibitem{Frezzotti:2003xj}
  R.~Frezzotti and G.~C.~Rossi,
  Nucl.\ Phys.\ Proc.\ Suppl.\  {\bf 128} (2004) 193
  [arXiv:hep-lat/0311008].

\bibitem{DellaMorte:2004wn}
M.~Della Morte,
Nucl.\ Phys.\ Proc.\ Suppl.\  {\bf 140} (2005) 458
[arXiv:hep-lat/0409012].

\bibitem{Sint:2005qz}
  S.~Sint,
  PoS {\bf LAT2005} (2005) 235
  [arXiv:hep-lat/0511034].

\bibitem{Frezzotti:2005zm}
  R.~Frezzotti and G.~Rossi,
  arXiv:hep-lat/0507030.

\bibitem{Frezzotti:2005gi}
  R.~Frezzotti, G.~Martinelli, M.~Papinutto and G.~C.~Rossi,
  arXiv:hep-lat/0503034.

\bibitem{Frezzotti:2003ni}
  R.~Frezzotti and G.~C.~Rossi,
  JHEP {\bf 0408} (2004) 007
  [arXiv:hep-lat/0306014].

\bibitem{Bernard:1987pr}
C.W.~Bernard, T.~Draper, G.~Hockney and A.~Soni,
Nucl.\ Phys.\ Proc.\ Suppl.\  {\bf 4} (1988) 483.

\bibitem{DellaMorte:2003mn} M.~Della Morte, S.~D\"urr, J.~Heitger,
H.~Molke, J.~Rolf, A.~Shindler and R.~Sommer [ALPHA Collaboration],
Phys.\ Lett.\ B {\bf 581} (2004) 93
[Erratum-ibid.\ B {\bf 612} (2005) 313]
[arXiv:hep-lat/0307021].

\bibitem{Borrelli:1992fy}
A.~Borrelli and C.~Pittori,
Nucl.\ Phys.\ B {\bf 385} (1992) 502.

\bibitem{Eichten:1989kb}
E.~Eichten and B.~Hill,
Phys.\ Lett.\ B {\bf 240} (1990) 193.

\bibitem{DiPierro:1998ty}
M.~Di Pierro and C.T.~Sachrajda  [UKQCD Collaboration],
Nucl.\ Phys.\ B {\bf 534} (1998) 373
[arXiv:hep-lat/9805028].

\bibitem{Palombi:2002gw}
F.~Palombi, R.~Petronzio and A.~Shindler,
Nucl.\ Phys.\ B {\bf 637}, 243 (2002)
[arXiv:hep-lat/0203002].

\bibitem{Bode:1999sm}
A.~Bode, P.~Weisz and U.~Wolff  [ALPHA collaboration],
Nucl.\ Phys.\ B {\bf 576} (2000) 517
[Erratum-ibid.\ B {\bf 600} (2001) 453; Erratum-ibid. B {\bf 608}
  (2001) 481] 
[arXiv:hep-lat/9911018].

\bibitem{Palombi:npren}
F.~Palombi, M.~Papinutto, C.~Pena and H.~Wittig [ALPHA collaboration],
work in progress.


\end{thebibliography}
